\documentclass[12pt]{article}
\usepackage{hyperref}
\hypersetup{
colorlinks=true,
linkcolor=black,
citecolor=black,
urlcolor=cyan,
filecolor=black,
pdfborder={0 0 0},
}

\usepackage{amssymb}
\usepackage{amsfonts}
\usepackage{mathrsfs}
\usepackage{latexsym}
\usepackage{amsmath}
\usepackage{graphics}
\usepackage{graphicx}
\usepackage{sectsty}
\usepackage{setspace}
\usepackage{multirow}
\sectionfont{\normalsize\bf}
\subsectionfont{\rm\normalsize}

\usepackage{tikz}
\usetikzlibrary{matrix,arrows,decorations.pathmorphing}

\usetikzlibrary{positioning}

\usepackage{wrapfig}
\usepackage{subfig}

\def\be{\begin{equation}}
\def\ee{\end{equation}}
\def\bd{\left|\begin{matrix}}
\def\ed{\end{matrix}\right|}
\input amssym.def
\input amssym
\def\aa{}

\def\Zop{{\Bbb Z}}
\def\M{{\cal M}}

\def\U{{\cal U}}

\def\L{{\cal L}}

\def\half{{\scriptstyle {1\over 2}}}

\def\Er{\hbox{E}}
\def\Ar{\hbox{A}}
\def\bphi{\boldsymbol{\phi}}
\def\br{\boldsymbol{r}}

\def\Rop{{\mathbb R}}
\def\bB{{\boldsymbol B}}

\def\bt{{\boldsymbol t}}
\def\SU{\hbox{SU}}
\def\SO{\hbox{SO}}
\def\U{\hbox{U}}
\def\G{{\cal G}}
\def\g{{\sf g}}

\def\s{{\sf s}}
\def\x{{\sf x}}
\def\y{{\sf y}}
\def\nox{{\scriptstyle{o \atop o}}}

\def\vac{|0\rangle}  
\def\bg{{\boldsymbol g}}
\def\bq{{\boldsymbol q}}
\def\balpha{{\boldsymbol\alpha}}
\def\blambda{{\boldsymbol\lambda}}

\makeatletter
   \def\tagform@#1{\maketag@@@{[#1]\@@italiccorr}}
\makeatother
\begin{document}

\pagenumbering{gobble}
\begin{figure}
  \includegraphics[width=\linewidth]{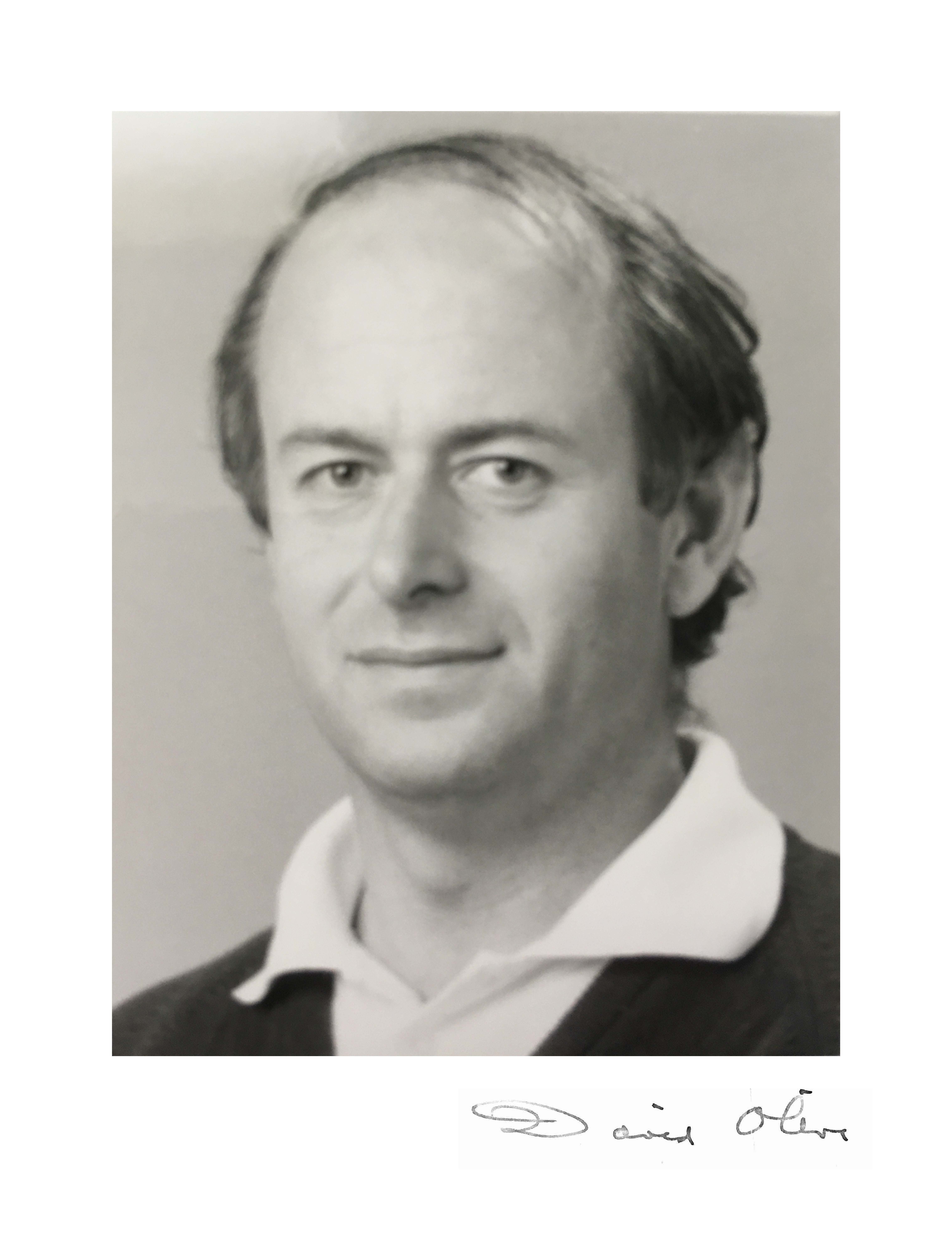}
\end{figure}

\vfil\eject

\thispagestyle{empty}
\baselineskip=16pt
\vspace{.5in}
{
\begin{center}
{\bf\large David Olive}
\vskip12pt
{\bf his life and work}
\end{center}}
\vskip 1.1cm
\begin{center}
{Edward Corrigan}
\vskip5pt

\centerline{\em Department of Mathematics, University of York, YO10 5DD, UK} 
\bigskip
\bigskip        
{Peter Goddard}
\vskip5pt

\centerline{\em Institute for Advanced Study, Princeton, NJ 08540, USA}
\centerline{\em St John's College, Cambridge, CB2 1TP, UK}
\bigskip
\bigskip
\bigskip
\bigskip
\end{center}

\centerline{\bf Abstract}

\bigskip
\bigskip
\leftskip=10truemm\rightskip=10truemm

David Olive, who died in Barton, Cambridgeshire, on 7 November 2012, aged 75, was a theoretical physicist who made seminal contributions to the development of string theory and to our understanding of the structure of quantum field theory. In early work on $S$-matrix theory, he helped to provide the conceptual framework within which string theory was initially formulated. His work, with Gliozzi and Scherk, on supersymmetry in string theory made possible the whole idea of superstrings, now understood as the natural framework for string theory. Olive's pioneering insights about the duality between electric and magnetic objects in gauge theories were way ahead of their time; it took two decades before his bold and courageous duality conjectures began to be understood. Although somewhat quiet and reserved, he took delight in the company of others, generously sharing his emerging understanding of new ideas with students and colleagues. He was widely influential, not only through the depth and vision of his original work, but also because  the clarity, simplicity and elegance of his expositions of new and difficult ideas and theories provided routes into emerging areas of research, both for students and for the theoretical physics community more generally.

\leftskip=0truemm\rightskip=0truemm
\bigskip\bigskip\bigskip
\bigskip
\bigskip
\begin{center}
[A version of section I {\sc Biography} is to be published\\ in the {\it Biographical Memoirs of Fellows of the Royal Society}.]

\end{center}
\setlength{\parindent}{0pt}
\setlength{\parskip}{6pt}

\setstretch{1.05}
\vfill\eject

\pagenumbering{arabic}

\bigskip

\setlength{\parindent}{0pt}
\setlength{\parskip}{6pt}

\setstretch{1.05}
\vfill\eject
\vskip50pt
\centerline{\large\sc I\quad Biography}
\bigskip\bigskip
\centerline{\sc Childhood}

\bigskip

David Olive was born on 16 April, 1937, somewhat prematurely, in a nursing home in Staines, near the family home in Scotts Avenue, Sunbury-on-Thames, Surrey. He was the only child of Lilian Emma (n\'ee Chambers, 1907-1992) and Ernest Edward Olive (1904-1944). Ernest worked as a clerk in the Bank of Belgium in the City of London. David believed the Olive family to be of Huguenot origin, although he was unable to establish this definitely. However the male line of the family could be traced back to the eighteenth century in the London area through his grandfather Thomas Henry Olive (1866-1948), an interior decorator, his great grandfather, William Henry Olive, a publican in Richmond, and William's father, James John Olive (1797-1869), a brazier and gas fitter. 

His maternal grandfather, Frederick McCall Chambers, was a music hall artist, who apparently disappeared mysteriously from the family scene. David had two cousins on his mother's side and eight on his father's but only one of them received a university education and there seems to be no scientific, or particularly intellectual, background in his family.

By 1940, David's family had moved to Walton-on-Thames, where he began his schooling at a local kindergarten. In 1941, to contribute to the war effort, Ernest resigned his post with the Bank of Belgium and volunteered for service in the Royal Air Force, which left the family in need of income. They moved to a bungalow in Govett Avenue, Shepperton, and Lilian took in a lodger and acquired a part-time job. David briefly attended Shepperton Grammar School before his mother decided in early 1943 that they should move to Edinburgh, partly to escape the bombing raids and partly to be near her mother. There, they lived in rented accommodation in Joppa, a suburb of Edinburgh, eventually taking rooms in the house of Mrs Susan Gage, the widow of the former steward of the Muirfield Golf Club. David struck up a friendship with Leslie, the younger of Mrs Gage's two sons, even though he was five years older than David. He joined Leslie at the Royal High Preparatory School and their friendship lasted a lifetime.

Late in 1943, Mrs Gage invited the Olives to move with her temporarily into the Muirfield Club House, where she had been invited to stand in for the steward. Ernest joined them there on leave over Christmas. This happy occasion was the last time that David saw his father, because, on 25 February 1944, Ernest died when the Lancaster bomber, on which he was serving as a flight engineer, was shot down over Germany. Lilian, understandably distraught, could not bring herself to give the news that Ernest was missing in action to the six-year-old David and eventually she left a letter announcing his father's death for David to read. Other disasters followed: the family furniture stored in London was destroyed in a fire, a loss not covered by insurance; a deposit that had been paid on a house being built was forfeited; and, in a particularly cruel stroke, no widow's pension was forthcoming from the Bank of Belgium because Ernest had resigned his position there to volunteer for the RAF, rather than waiting to be conscripted. 

Soon after the loss of her husband, Lilian decided to return to England, to Birchington in Kent, where David attended Woodford House School, but the move did not last. To get away from the V-1 and V-2 rocket bombs, and also because they found the education at Woodford House inferior to that provided by the Royal High School, David and his mother went back to Scotland in the spring of 1945. Lilian bought a bungalow in Newhailes Avenue, Musselburgh, near Edinburgh, next to friends, Jimmie and Mabel Weatherhead, parents of one of David's classmates. Jimmie, a local bank manager, was in a position to help, at least with financial advice.  

The Royal High School had the further advantage that no fees would be charged provided that David performed sufficiently well academically. This was a condition he had no difficulty in meeting, ending up top of his class every year from 1946 onwards. He left the junior part of the Royal High School in 1949 and, subsequently, in 1955, the senior part, as {\it Dux} (the leading academic student). He remembered two of the women who taught him as really inspiring: Hilary Spurgeon, who gave David his first science courses, and Letitia Whiteside, who provided extra science classes after hours. But the others were, in his view, lacklustre: the man who taught physics just read from a text book, while the head of science was `awful' (although, it seems he added some colour by standing on a stool at the end of the last lesson of the day and proclaiming, after Horace,  {\it Odi profanum vulgus}).

For all this, it was not at school that David found the direction of his future scientific career. 
\begin{wrapfigure}{r}{65mm}
\vskip-6pt
\centering
  \includegraphics[width=60mm]{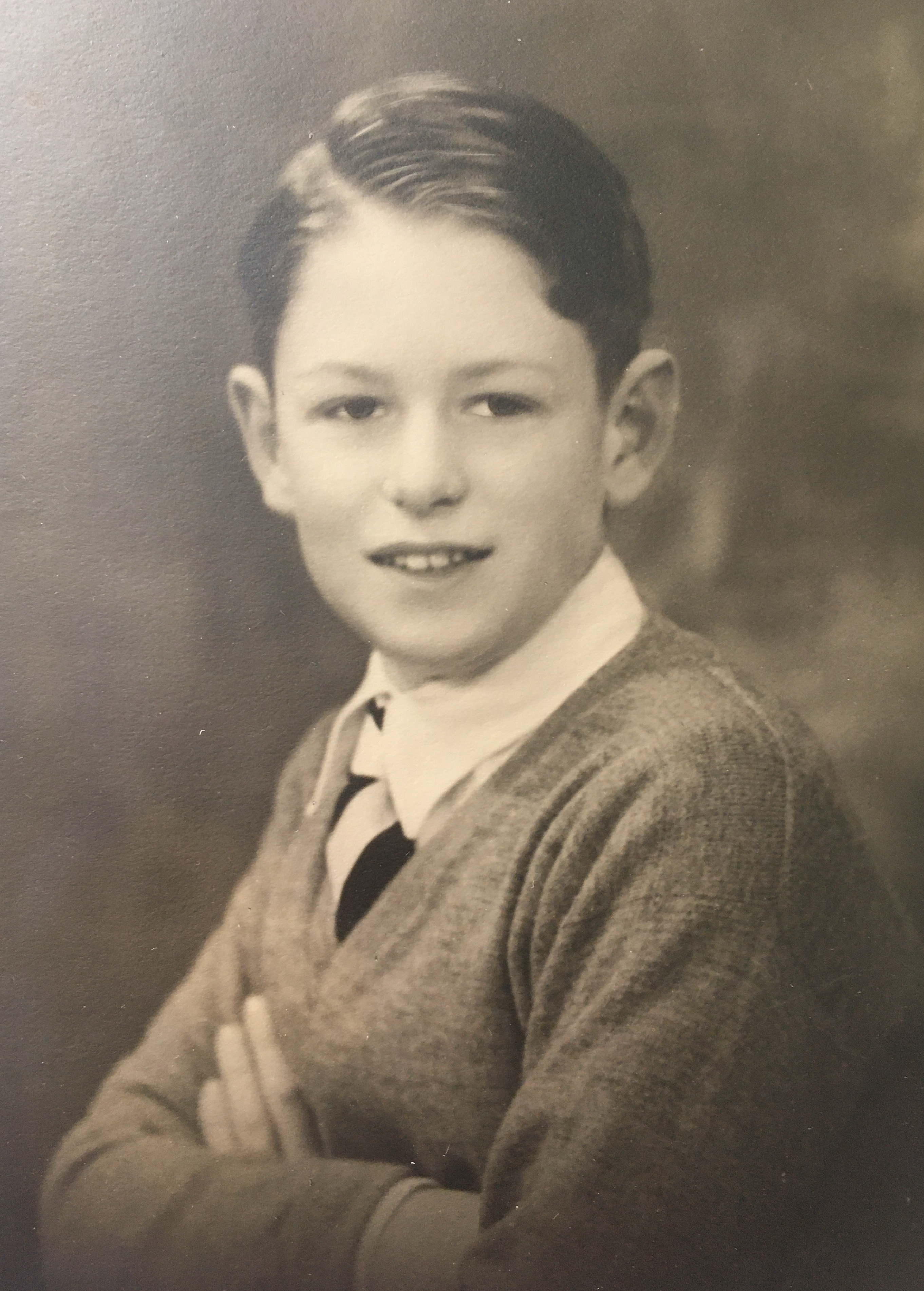}

\vskip2pt  
\scriptsize David Olive aged about 11. \\
Photograph
kindly provided by the Olive family.
\vskip-18pt
\end{wrapfigure}
In David's view, more important than anything he learnt at school was the time he spent 
with his Meccano set, building its little metal `girders', together with nuts, bolts, wheels and gears into relatively complicated mechanical devices in the form of cranes, lorries, cars, etc., sometimes powered by electric motors. The more complicated, the longer was the time that might be taken for the construction, up to months in some cases. David felt that this encouraged him later, in his research career, to conceptualize long-term programmes, aimed at achieving complicated objectives. 

As a teenager, David developed other interests that would stay with him throughout his life. Beginning in 1952, Lilian encouraged David to take an interest in golf. Perhaps the seed had also been sown by that last family Christmas with his father at Muirfield. He joined the local club as a junior member, had lessons with the club professional, and became secretary of its junior section. Inspite of his love of the game, in his own estimation, David was never very distinguished as a golfer, but a number of his contemporaries at the club became quite well known. 

An even more consuming interest, music, which was to become a life-long passion, had its beginnings in these years. This was encouraged by his friend, Leslie Gage, who had a large radiogram. Starting with Beethoven, David soon progressed to Mozart, Berlioz, Delius and many others, both the widely familiar and others rarely heard. He became a great admirer of the conductor Sir Thomas Beecham and, at one Edinburgh Festival, he and Leslie Gage managed to slip into the Usher Hall to sit in on some of the great man's rehearsals. 

David left the Royal High School laden with academic honours, including bursaries and scholarships for Edinburgh University. In his farewell speech as Dux of the school, foreshadowing his future intellectual path, he chose to talk about theoretical physics, a subject which already fascinated him. 

\bigskip
\bigskip
\bigskip

\centerline{\sc Undergraduate Years}

\bigskip

After matriculating at Edinburgh University in September 1955, David continued to live at home. He began by taking the courses in Mathematics, Mathematical Physics and Natural Philosophy (as physics was then known there). The Mathematical Physics lectures took place in the Tait Institute, newly opened in 1955, which had been established for Nicholas Kemmer. Appointed as Tait Professor in 1953, initially within the Natural Philosophy (physics) department, Kemmer had found that Norman Feather, the holder of the more ancient chair of Natural Philosophy, thought that there was no such thing as theoretical physics, because, in his view, physics was in essence experimental. So, Kemmer had secured an independent institute for theoretical physics. Against this background, when David had to choose two of the three courses he was studying to specialize in for his degree, he decided to drop physics, rather than mathematics or mathematical physics, even though he continued to attend physics lectures throughout his second year. 

David found Kemmer's lectures on hydrodynamics, given without reference to notes, wonderfully, indeed enviably, clear. He also attended lectures by John Polkinghorne, whose first lecturing post (1956-8) was in the Tait Institute, and who was later to become David's colleague and collaborator in Cambridge. He continued to excel academically, as he had at school, only held back at times by his nearly illegible handwriting, which caused at least one examiner to mark him down.

A large part of David's social life centred on the Edinburgh University Physical Society, which organized hiking expeditions in addition to lectures on physics. Through the society and other activities, he met many of his future colleagues and friends, such as Keith Moffatt, David Fairlie, Jim Mirrlees, Ian Drummond, Ian Halliday, Tom Kibble, and Alan MacFarlane, all exact or near contemporaries. They were just part of a remarkable succession of students in theoretical physics and mathematics at Edinburgh at that time, and Kemmer's charismatic influence inspired many of them to follow research careers in fundamental physics.  

A lecturer who sometimes joined the physical society hikes was William Edge, whose specialty was projective geometry. Edge, who was perhaps not completely at home in the twentieth century, encouraged many of the more talented students, including David, to go to Cambridge after graduation from Edinburgh, to follow a path, well-trodden at least since the time of James Clerk Maxwell, of reading the Mathematical Tripos for a second undergraduate degree. In December 1957, before he finished his Edinburgh degree, which he completed in three years rather than the usual four, obtaining first class honours, David competed successfully for an open scholarship at St John's College, Cambridge, on Edge's advice.

After Edinburgh, David spent the summer working at Metropolitan Vickers, an electrical engineering company in Manchester, and it was from there he put his trunk on the train to go up to Cambridge in October, 1958. As a Scholar of St John's, he was given rooms in College, at C2 Chapel Court, in a part of the College with the relatively modern convenience of a toilet on each staircase, although it was still necessary to cross the court in order to get a bath. In his first year, he studied for Part II of the Mathematical Tripos, then taken by students coming to Cambridge directly from school in either their second or, more commonly, third year, and he again obtained first class honours. Having experienced the intensity of the Cambridge course, he came to feel that he had benefited from the somewhat slower pace of the Scottish system, and he said that this led him to think of himself as Scottish, in spite of having been teased at school for being a `sassenach'. 

At some point during his first term, he attended a meeting of the Cambridge University Heretics Society, at which, perhaps appropriately enough, the speakers did not show up. David and some other members of the society decided to go out to a pub instead. In this way, David met Jenny Tutton, then in her second year reading mathematics at Girton College, whom he was to marry four and a half years later. 

In his second year at St John's, in order to complete the requirements for the BA degree, which Cambridge allows graduates of other universities to take in two years rather than three, 
David studied for Part III of the Mathematical Tripos, taking courses by John Ziman on Solid 
State Physics, Christopher Zeeman on Algebraic Topology, Fred Hoyle on General Relativity and Cosmology, among others. The courses David found most inspiring included one on Quantum Field Theory by John Polkinghorne, who had returned to Cambridge from Edinburgh in 1958, and, especially, two courses on Quantum Mechanics by Paul Dirac, whose expositions and contributions to physics were to have a very profound influence on David's approach to research in theoretical physics. The Part III examinations at the end of the academic year took place at the height of the hay fever season and, perhaps because of this, although David passed, he failed to be awarded a distinction, the mark necessary for an assured place to remain in Cambridge to undertake research for a PhD. Nevertheless, his special talents had been recognized, the Science Research Council awarded him a research studentship and, not withstanding his somewhat disappointing examination performance, he was allowed to stay on. 

David spent his last summer before starting research in Austria. He had obtained a grant from the Austrian Institute to attend a German course at the University of Vienna, which lasted for most of July, but David stayed on until late September. His interest in music had not diminished at all during his undergraduate years and here he had the leisure to attend many performances at the Vienna State Opera, including the whole {\it Ring} cycle and {\it Tristan und Isolde}, conducted by Herbert von Karajan, {\it Der Rosenkavalier}, {\it Capriccio} and (at the Redoutensaal) {\it Cosi Fan Tutte}, conducted by Karl Bohm, as well as {\it Aida}, conducted by Lovro Von Matacic, as he recorded in the very detailed notes that he kept on the concerts and opera performances he attended over the years. 
\bigskip
\bigskip
\bigskip

\centerline{\sc Beginning Research at Cambridge}
\nopagebreak
\bigskip

David began research in October 1960, working within the Department of Applied Mathematics and Theoretical Physics (DAMTP) in Cambridge, which had been established just a year earlier. 
\begin{wrapfigure}{r}{85mm}
\vskip-6pt
\centering
  \includegraphics[width=80mm]{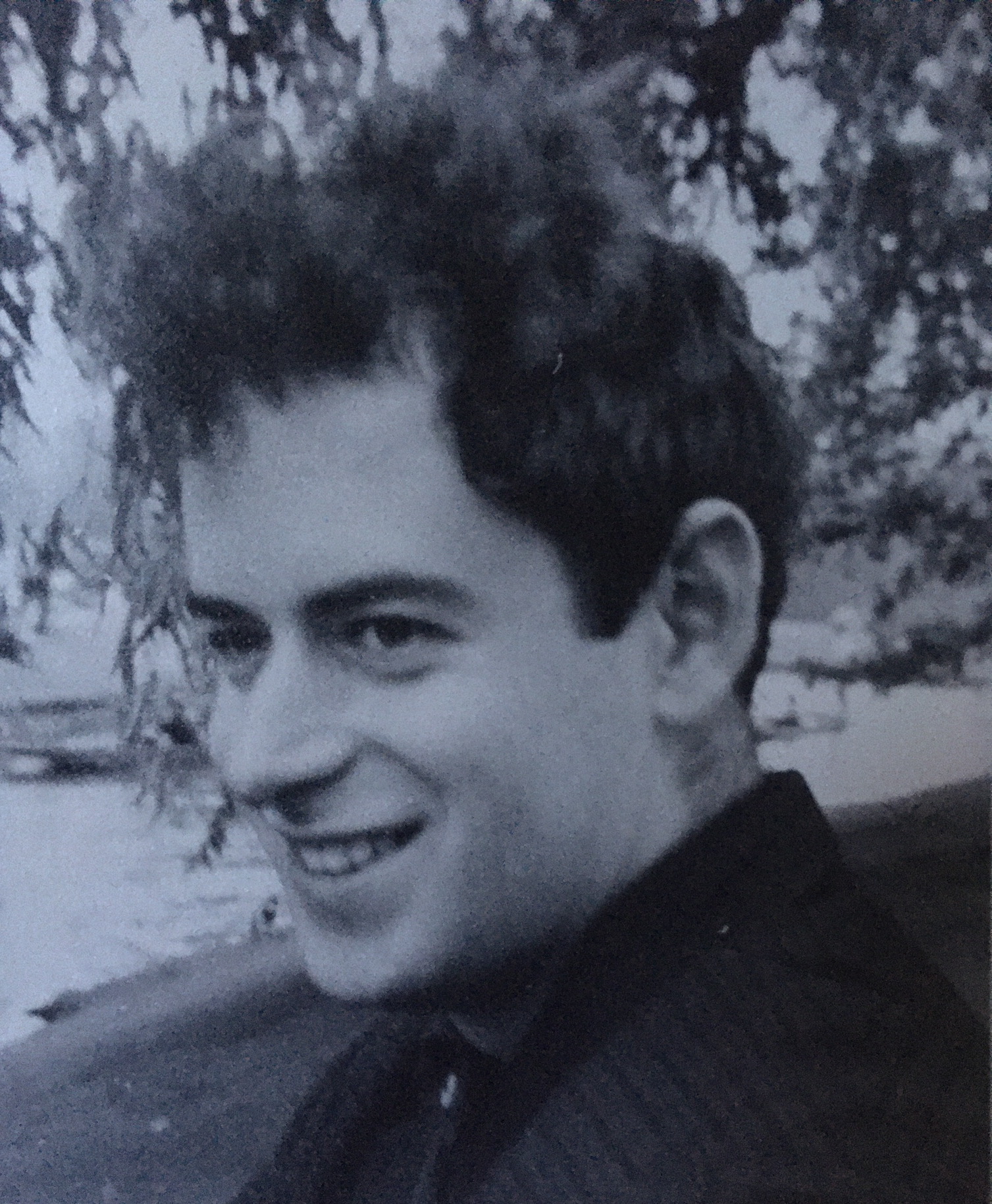}

\vskip2pt  
\scriptsize David Olive as a student in Cambridge. \\
Photograph
kindly provided by the Olive family.
\vskip-18pt
 \end{wrapfigure}
At first, David had lodgings in Park Parade, but his landlady objected to his habit of playing 
classical music loudly and he moved to Alpha Road at the end of the term. One term later, he joined DHJ (Ben) Garling and Johnson (Joe) Cann, fellow graduate students at St John's College, in a flat in Newnham Road. Both Ben and Joe found David to be very undomesticated; his mother had looked after him so well that he had acquired no practical skills in cooking or housekeeping.

The particle theory research students in DAMTP were housed in the Austin Wing of the old Cavendish Laboratory off Free School Lane in the centre of Cambridge. His contemporaries then included David Bailin and Ian Drummond and, in the year ahead of him, Peter Landshoff. John C Taylor, who had just returned to Cambridge from Imperial College, was appointed as David's research supervisor. Weekly seminars were attended by Paul Dirac, then Lucasian Professor of Mathematics, as well as the other faculty: Richard Eden, John Polkinghorne and John C Taylor.

Taylor initially suggested to David that he try to find a renormalizable theory of the weak interactions based on Yang-Mills gauge theory. This was an extremely ambitious objective, eventually successfully attained through the work of Glashow, Salam and Weinberg, and of 't Hooft and Veltman (for which they received Nobel Prizes in 1979 and 1999, respectively). Convinced by an earlier paper by Salam and Komar (1960) that it was impossible to construct a renormalizable theory of massive gauge particles, David's interest shifted to what was then the local speciality of studying the analytic properties of perturbative quantum field theory. His first paper (1), written in collaboration with JC Taylor, early in his second year as a research student, was a contribution to the understanding of the complicated singularity structures that can occur, in particular acnodes and cusps. 

From this work, his interest developed towards the then current attempts to formulate an axiomatic theory of the scattering matrix, known as the $S$-matrix, and this was the first area in which David would make important contributions. The leading proponent internationally of this approach was Geoffrey Chew of the University of California at Berkeley, who fortuitously spent the academic year 1962-63 in Cambridge as a Visiting Fellow of Churchill College. 

On 15 April 1963, David married Jenny Tutton at The Catholic Church of Our Lady, Belper, Derbyshire, with Joe Cann acting as best man. After a honeymoon in Paris, David and Jenny returned to an upstairs flat in Devonshire Road, near the railway station in Cambridge, convenient for Jenny's daily journey to work as a mathematics teacher at the Hertfordshire and Essex County High School for Girls in Bishop's Stortford. [Jenny's two younger sisters, Rodie and Clare, were also to marry Cambridge theoretical physicists: Rodie married Tony Sudbery, who, after taking his PhD in Cambridge, spent his career in the University of York, and Clare married Ian Drummond, who was to be David's faculty colleague in DAMTP.]

Just before his PhD examination, conducted by Gordon Screaton and (following the practice then usual in Cambridge) his supervisor, JC Taylor, David was elected to a Research (i.e. Postdoctoral) Fellowship at Churchill College, along with Ian Drummond, possibly (as David thought) thanks to the backing of Chew. His doctoral dissertation was entitled {\it Unitarity and $S$-matrix Theory}.

On September 7th, David and Jenny set sail on the Berlin from Southampton to New York to travel to Pittsburgh, to spend the academic year at Carnegie Tech (now Carnegie Mellon University) at the invitation of Dick Cutkosky. It was to be David's only year away from Cambridge as a postdoctoral fellow. Whilst he was there, David restructured the unpublished paper,  {\it Towards an axiomatization of $S$-matrix theory}, which he had written just before leaving Cambridge, and published it as {\it An exploration of $S$-matrix theory} (5), an approach to a self-consistent determination of the singularity structure of the $S$-matrix.
As well as the research life in the physics department and social contacts with its members, the Olives enjoyed the concerts available in Pittsburgh and David took the opportunity to acquire new amplifiers, speakers and turntable for his audio system. However, when Jenny became pregnant with their first child, they decided to return to Cambridge, where David's fellowship at Churchill College would provide accommodation and a secure salary for a few years.

Arriving back in Cambridge in August 1964, David and Jenny moved into a newly built flat at Churchill College. David took part in the social and intellectual life of the College, presided over by its founding Master, Sir John Cockroft, until his death in 1967. Cockcroft attracted many distinguished physicists to the College as visitors, including Peter Kapitza, Mark Oliphant, George Gamow and Murray Gell-Mann. 

On December 5th, 1964, Jenny gave birth by Caesarean section to a daughter, Katie, who weighed less than 3 pounds and spent some weeks in an incubator. With the family growing, and David appointed to an Assistant Lectureship in the University of Cambridge from October 1965, he and Jenny purchased a house with a sizable garden in the village of Barton, about 5 miles from Cambridge, which they continued to own for the rest of their lives and to which they returned in retirement. 
Earlier, in the first months of 1965, David gave a lecture course, his first, a graduate course on $S$-matrix theory, based on the paper he had written in Pittsburgh (5). These lectures in turn became the basis for David's contribution to the book, {\it The Analytic $S$-Matrix} (12), written together with his Cambridge colleagues, Richard Eden, Peter Landshoff and John Polkinghorne. 
Completed against a tight timetable in July 1965, and published by Cambridge University Press the following year, the book's four chapters were each assigned principally to one of the authors, with David contributing the final chapter, $S$-matrix theory. Exceptionally, for a book on particle 
physics, {\it The Analytic $S$-matrix}, known affectionately by the initials of its authors, ELOP, has remained in active use as the standard reference on the subject. The first three chapters comprise a general introduction, a discussion of the singularity structure of the Feynman graphs describing the perturbative treatment of quantum field theory, and an account of methods for analyzing the high-energy behavior of Feynman graphs, these two topics being ones to which their respective authors had made important contributions, as David had to $S$-matrix theory. Characteristically, rather than rely on the material of his co-authors, David's treatment in the fourth chapter begins practically {\it de novo}, and so can be read independently of the rest of the book. 

Throughout his career, nearly all of David's research was done in collaboration, often working with a particular colleague periodically for over a decade or more, but he always sought to build up his own understanding of a topic from a set of basic principles or assumptions, which he would analyze and simplify, repeatedly going through the arguments to find the simplest, most elegant and rigorous discussion of the topic. 

In 1966, David was promoted to a Lectureship in the University, an effectively tenured position, and, on 25 October, Jenny gave birth to their second child, Rosalind. He continued to work on $S$-matrix theory and extended his interest to encompass the Regge theory of the high energy behavior of scattering amplitudes. 

In August 1968, he travelled again to Vienna to attend the International Conference on High Energy Physics, the fourteenth in a series of conferences, then held every two years, which brought together leading experimentalists and theoreticians from around the world. As usual, he took full 
advantage of what was available musically while he was there, including a performance of {\it The Magic Flute}, with Dietrich Fischer-Dieskau, and a Franco Zeffirelli production of {\it La Boh\`eme}, with Renate Holm and Cesare Siepi. But, as he noted (131), it was an unexpected talk given in the marbled ballroom of the Hofburg Palace that was, despite the poor acoustics of the conference location, to change the direction of his research and, soon after, of his life. 

The talk, by Gabriele Veneziano, then not quite 26 years old, introduced his soon-to-be famous formula for a two particle scattering amplitude. Veneziano's objective was to illustrate how analytic properties could be combined with Regge asymptotic behavior in an explicit mathematical function (Veneziano 1968), but it quickly stimulated attempts at generalization and eventually proved to be the seed that grew into string theory, one of the most influential developments in fundamental physics in the twentieth century, and one in which David was to play a leading role. 

Back in Cambridge, David gave a general talk in DAMTP on Veneziano's work, and remembered being taken aback by Dennis Sciama's prescient suggestion that this might be the start of a new theory. He began collaborating with David Campbell and Wojtek Zakrzewski, then research students, on finding extensions of Veneziano's formula to the scattering of particles with spin (18--20). While this work did not find a permanent place in the development of the subject, it served to focus David's interest.

\bigskip
\bigskip
\bigskip

\centerline{\sc Moving to CERN}

\bigskip

Early in 1969, David gave a seminar at CERN, Geneva, at the invitation of Andr\'e Martin, and this motivated David to apply to spend his upcoming term of sabbatical leave there. In September, David and his family began a three-month stay, renting a CERN apartment in Rue du Livron, Meyrin, and taking the opportunity to get to know the Swiss countryside, through visits to Chillon, Gruy\`eres, Lauterbrunnen and, particularly, Annecy, nearby in France.

He soon met Daniele Amati, then a CERN staff member, who introduced him to Michel Le Bellac and suggested that the three of them collaborate. Amati was an Italian physicist, with great charisma, who had grown up in Argentina: he had a poster of Che Guevara on his office door, and drove a Bentley, which he never locked and readily loaned to David and others in temporary need of a car.  Amati had a great talent for stimulating lively research discussions and for encouraging younger physicists, a talent strangely lacking in some of the other staff members in the CERN Theory Division at the time. He was to have a major influence on David. 

David wrote two papers (23, 24) with Amati and Le Bellac on the operator formalism for dual models, as the development of Veneziano's breakthrough had become known. Amati arranged for David's stay at CERN to be extended by a further three months at the beginning of 1970, for which David took unpaid leave from Cambridge. This time Jenny and their daughters stayed in Cambridge. David took a room in Chemin du Vieux Bureau, Meyrin, and devoted his leisure time to skiing and, as always, music. A recital of Beethoven sonatas by Wilhelm Kempff captivated him and he remained a devotee of the great German pianist for the rest of his life.

The collaboration with Amati and Le Bellac was joined by Victor Alessandrini, a younger Argentinian physicist, and the four agreed to give a series of six lectures on dual models, which then became the basis for an influential review article, {\it The operator approach to dual multiparticle theory} (26). When he returned to Cambridge, David gave a version of these lectures in DAMTP during May, which were attended by Ed Corrigan, Peter Goddard, Michael Green and, probably, Jeffrey Goldstone, all of whom went on to make substantial contributions to the subject.

During David's second term at CERN, Amati raised with him the possibility of his spending a longer time there as a staff member, for three years in the first instance. There was no possibility of obtaining leave from Cambridge for such an extended period under the University's policies at the time. John Polkinghorne, the leader of the theoretical physics group within DAMTP, made it clear to David that it might be possible for him to return to a faculty position there, but that was by no means guaranteed. As David later put it, he was making an enormous gamble, because he would be giving up a tenured post in Cambridge for a fixed-term one at CERN; but, he was prepared to make this sacrifice, in order to be able to spend all his time on the theory of dual models, because he thought this might well be the theory of the future.

At the end of June 1971, David left Cambridge to take up what was eventually to be a six-year staff position at CERN. For the first month, David stayed in the hostel on the top floor of the building housing the CERN Theory Division. He quickly met other new arrivals with interests in physics similar to his own, including Lars Brink from Gothenburg, who was just beginning a postdoctoral fellowship and who was to become one of David's collaborators and a life-long friend.  He was joining the group of mainly young theoretical physicists working on dual models that had gathered around Daniele Amati, possibly the largest group in the world working on the subject. It was big enough to sustain a weekly seminar, meeting on Thursdays at 2 pm, with those attending regularly at times during David's first year at CERN including Alessandrini, Amati, Brink, Corrigan, Di Vecchia, Frampton, Goddard, Rebbi, Scherk and Thorn. 

The atmosphere was informal and collaborative, with a very free exchange of ideas. There was the feeling of participation in a shared enterprise to construct a new and radically different theory, as well as a camaraderie engendered by the active disapproval of many of the senior physicists at CERN and elsewhere. It seemed possible that a fully consistent theory of the strong interactions might be fashioned out of dual models, with the very requirement of consistency narrowing down the range of dual models that should be considered as physically relevant. There was the sense that the theory was thus defining itself, rather being crafted by those working on it, and it was exciting to see its form, different from quantum field theory, emerge before one's eyes, from within itself.  David's background in $S$-matrix theory, which provided the conceptual context within which the theory could be defined, together with his ability to find precise, elegant and simple arguments, made it an ideal research area for him.

When David's family arrived, they moved into a spacious, newly built apartment on the ninth floor at Le Lignon, with expansive views towards the Sal\`eve and the Alps beyond. David was to find living in Geneva somewhat of a culture shock for someone brought up in Edinburgh, though a shock that was not completely unwelcome. At a mundane level, David found it a relief not to have his Barton garden to tend and, without the teaching and administrative duties he had had in Cambridge, he was free to concentrate on research, while also having time for his usual leisure pursuits. He joined the CERN golf club and he purchased a season ticket for the opera and ballet season at the Grand Th\'e\^atre in Geneva. 

At weekends the family frequently visited the mountains, hiking in summer and skiing in winter, often in the nearby Jura mountains.  While Jenny and Katie preferred cross country skiing, David took Rosalind downhill skiing, where he struck a characteristically nice balance between encouraging adventure and providing reassurance and caution when appropriate. David remained a keen skier past his years at CERN, taking advantage of the opportunities provided by conferences in Les Houches and other mountain resorts whenever he could.

Ed Corrigan, then David's research student, came to CERN for two months in late 1971, and they worked together on building a consistent dual model that included fermions. At the beginning of the year, Pierre Ramond had introduced fermions into the theory in a way that looked extremely promising (Ramond 1971), and so it was to turn out, but many steps would be necessary to ensure that a full theory could be constructed that did not harbour inconsistencies. Most of David's efforts in the first half of his six-year tenure at CERN were directed towards this end, working mainly with Lars Brink. 

In 1972, through work at CERN and elsewhere, definitive progress was made on determining the physical states occurring in dual models, leading to an understanding of how the picture of dual models as describing the scattering of one-dimensional objects, referred to at the time as `rubber bands', `threads'  or `strings', suggested two or three years earlier by Nambu (1969), Nielsen (1969) and Susskind (1970), could be made precise in terms of the quantum theory of a what was now termed a `relativistic string' (Goddard {\it et al.} 1973). For some theoretical physicists, who had hoped that dual models would correspond to a more radically different sort of physical system, one that could not be given such a space-time interpretation, this was actually a disappointment, but, for David, who had followed these developments closely, it was a stimulating breakthrough and, he presented a brief account of it (29) at the 16th International Conference on High Energy Physics held at Fermilab, Batavia, Illinois, in the summer of 1972.

Having a detailed description of the physical states of the dual model enabled a precise formulation of the objective of constructing scattering amplitudes in the fermion theory as well as doing the same for the loop contributions that it was necessary to add into the original bosonic dual model of Veneziano. David and Lars Brink constructed the one-loop bosonic loop first (30,31), realizing this would be a useful technical preparation for calculating fermion amplitudes as well as an important step in itself. After this, David and Lars began a series of papers (32--34) on the fermion theory and other aspects of the physical states in dual theory, with Jo\"el Scherk, who had come to CERN from the \'Ecole Normale Sup\'erieure in Paris, joining the collaboration. This led in October 1973 to David's calculation, in collaboration with Corrigan, Goddard and Russell Smith, of fermion-anti-fermion scattering (36).  

Even as this progress was being made, and while the fascination and promise of dual models, or string theory, as the subject increasingly was being described, remained compelling in the eyes of those working on it, the interests of most physicists was being captured by the `standard model' of particle physics then being developed. In December, David Olive wrote to Peter Goddard,``Very few people are now interested in dual theories here in CERN. Amati and Fubini independently made statements to the effect that dual theory is now the most exciting theory that they have seen but that it is too difficult for them to work with. The main excitement [is] the renormalization group and asymptotic freedom, which are indeed interesting.'' It was becoming apparent that the gamble that David had taken in relinquishing tenure at Cambridge to take the staff position at CERN was a real one. As Paolo Di Vecchia has commented, ``we were so attracted to the beautiful properties of string theory that it seemed a waste to abandon it, but it was clear that, if we had continued [to work on it] for much longer, we would not have been able to get a permanent job.'' Indeed, David remembered Amati later warning him that ``you are unemployable because you do string theory''.

Against this background, the months after the completion of the fermion calculation were a fallow period for David's research, but he was invited to give a plenary session talk on progress on dual 
\begin{wrapfigure}{r}{105mm}
\vskip-6pt
\centering
  \includegraphics[width=100mm]{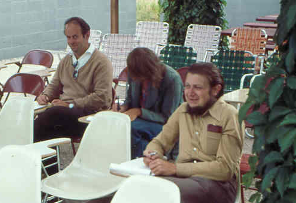}

\vskip2pt  
\scriptsize David Olive, at a workshop at the Aspen Center for Physics in Colorado, in August, 1974, seated next to Jo\"el Scherk (with head bowed) and with Eug\`ene Cremmer in the foreground. \\
Photograph
kindly provided by Lars Brink.
\vskip-18pt
\end{wrapfigure}
models at the 17th International Conference on High Energy Physics, held at Imperial College, London, at the beginning of July 1974. In his review (38), David sought not only to emphasize the conceptual formulation of dual models as a quantum theory of strings but also that dual (or string) theory contained within it electrodynamics, Yang-Mills gauge theory and Einstein gravity. Indeed, as he stressed at the beginning of his talk, consistency demanded that the theory contain massless spin 2, spin 1 and spin $\half$ particles, corresponding to the graviton, photon, and the neutrino, as seen in nature. Dual theory, conceived as a theory of strong interactions, ironically had produced from within itself the emerging theories of all the other fundamental interactions, offering the prospect of a unified theory.

\bigskip
\bigskip
\bigskip

\centerline{\sc An Apparent Change of Direction}

\bigskip
The London conference of 1974 was a turning point in David's research, not only because of the impact of his review talk, but even more as a result of a talk given by Gerard 't Hooft on magnetic monopoles in gauge theories ('t Hooft 1974). David later said he did not really understand what 't Hooft had done until he heard Murray Gell-Mann discuss it during a meeting at the Aspen Center for Physics, Colorado, which took place just after the London conference. This meeting, organized by John Schwarz, brought together many of those then working on string theory for what seemed in retrospect to be just about the last hurrah of the early years of string theory, before it was eclipsed by the rise of QCD and the standard model. Although, outwardly, the focus of his interests moved on to other subjects, like a number of the early dual model 
devotees, David's fascination with string theory never left him, and the major contributions that he made during the remainder of his career, even when they appeared unrelated to string theory, ended up playing a central role in its development down to the present day. 

After David returned to Geneva in the autumn of 1974, he began trying to understand the 't Hooft-Polyakov monopole,  as it came to be known following a paper from Alexander Polyakov (1974), which appeared at roughly the same time and covered similar ground to that of 't Hooft. Indeed, David spent most of his remaining three years at CERN elucidating the structure of monopoles in gauge theories, working initially with Ed Corrigan, who had come back to CERN as a postdoctoral fellow for two years, and with David Fairlie and Jean Nuyts, who joined the collaboration in 1975 (40, 42).

Nuyts had a familiarity with the theory of Lie algebras, and their roots and weights, and, in the summer of 1976, he and David began to realize that this part of mathematics provided the appropriate mathematical language for describing magnetic monopoles in general non-abelian gauge theories. Peter Goddard also came to CERN, as a visitor for the summer of 1976, and their discussions led to the paper, {\it Gauge Theories and Magnetic Charge} (46), which has had a long-term and continuing influence. The magnetic monopoles Goddard, Nuyts and Olive classified became known as `GNO' monopoles. 

David later recalled how it was while driving with his family to spend some days in Wengen in the Bernese Oberland that the key idea came to him. Just as the electric charges were associated with points on the weight lattice of symmetry group, the magnetic charges were associated with another lattice, the lattice dual to the weight lattice. David realized that this dual lattice was itself the weight lattice of another Lie group, which GNO called the `dual group'.  In the spring of 1977, Goddard met Michael Atiyah at a conference on mathematical education in Nottingham. Atiyah was becoming interested in theoretical physics, though he was not yet familiar with the then recent developments on monopoles in gauge theories. He immediately realized that the GNO dual group was the same as the dual group introduced by Robert Langlands (1970) within the context of what was known as the Langlands program, one of the major developments in pure mathematics in the second half of the twentieth century. However, it was nearly thirty years before the relationship between electric-magnetic duality in gauge theories and Langlands duality in the theory of automorphic forms began to be understood in depth.   
    
The magnetic monopoles identified by 't Hooft and Polyakov, and the generalizations studied by David and his collaborators, are extended objects in a gauge theory, and so have a different status a priori from the electrically charged particles, which are quanta of the fundamental fields in the theory. David formed the bold and prescient vision that there should be a dual formulation of the theory in which magnetic and electric charges and fields interchanged their roles. 

When Claus Montonen, who had been David's graduate student for a year before he left Cambridge, visited CERN in the spring and summer of 1977, David began working with him to find evidence for his duality conjecture. Together they considered the simplest theory containing 't Hooft-Polyakov monopoles and showed that the spectrum of magnetic charges of the monopoles was just like that of the electric charges of the fundamental quantum particles of the theory 
\begin{wrapfigure}{l}{105mm}
\vskip-6pt
\centering
  \includegraphics[width=100mm]{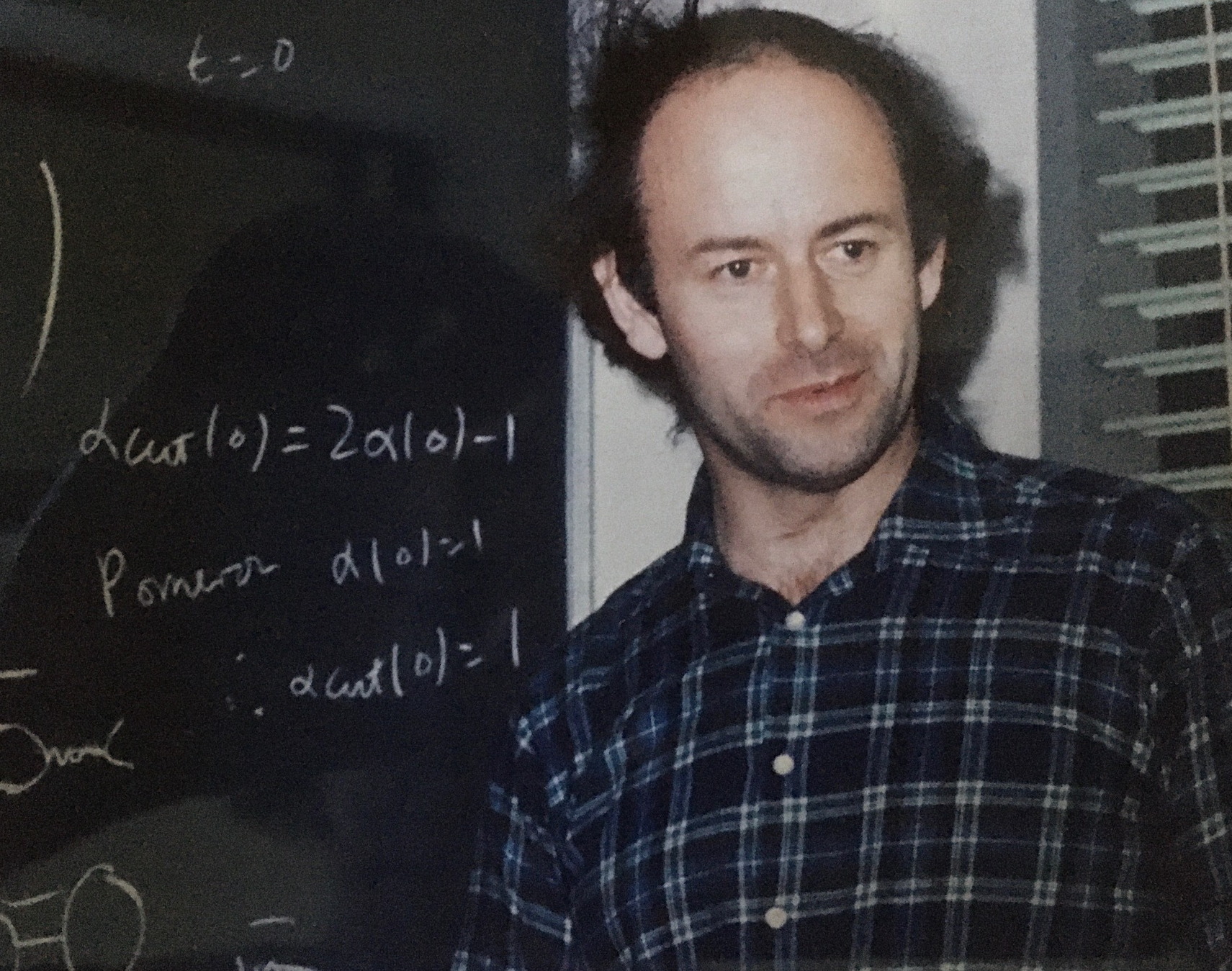}

\vskip2pt  
\scriptsize David Olive in Kyoto in April 1976. \\
Photograph
kindly provided by the Olive family.
\vskip-12pt
\end{wrapfigure}
and, moreover, the mass of the monopoles was given in terms of the mass and electric charge of these fundamental particles by exactly the same expression as that relating the fundamental particle mass to the monopole mass and magnetic charge. All this provides evidence for a duality symmetry between electrically charged fundamental particles and magnetic monopoles, which are extended objects in the theory. The electrically charged particles are the massive gauge bosons produced by the Higgs mechanism in the spontaneously broken theory, so Montonen and Olive 
conjectured that there was a dual formulation of the theory in which the magnetic monopoles became fundamental particles, the massive gauge bosons associated with a spontaneously broken dual gauge symmetry. 

Montonen and Olive refined their ideas over the months preceding David's departure from CERN in September 1977, but they felt that several problems remained. Characteristically, David was reluctant to write a paper before he felt that the arguments had reached their most elegant and succinct form. However, they were both convinced that they were on to something, and they drafted a paper, {\it Magnetic monopoles as gauge particles?} (47), which became one of David's most famous and influential research contributions, and the proposal that it put forward, of the existence of a dual magnetic formulation of a gauge theory, became known as Montonen-Olive duality or the Montonen-Olive conjecture. The formulation of this conjecture exemplified very well David's extraordinary ability to find precise, elegant and deep relations, encapsulating the essence of a physical situation, and build on them a bold imaginative vision of what further structures might await discovery.

\bigskip
\bigskip
\bigskip

\centerline{\sc Last Years at CERN}
\nopagebreak
\bigskip
Although the study of monopoles dominated David's last three years at CERN, his collaboration with J\"oel Scherk drew him again to dual theory for a significant interlude in 1976. At the \'Ecole Normale Sup\'erieure, Scherk had begun collaborating with a visitor, Fernando Gliozzi, and they had noticed that, after making a certain projection in the fermion dual model, which both removed something like half the states, and solved one of the model's problems by eliminating the unwanted tachyon (faster than light particle), the number of fermion states at any given mass equalled the number of boson states, which strongly suggested that the theory was spacetime supersymmetric. This projection as a means of removing the tachyon had been discussed informally since 1974 but it had not been realized that it resulted in an equal number of fermion and bosons states. On a visit to CERN to give a talk, Scherk told David about their results. David, who had earlier studied the properties of the Dirac equation in various dimensions of space-time, pointed out to them that the consistency of their projection required this dimension to differ from 2 by a multiple of 8, which fortunately included the space-time dimension 10, which was needed for consistency of the theory for other reasons. 

Building on their earlier work together, Scherk invited David to join the collaboration and together Gliozzi, Scherk and Olive wrote two papers on these ideas (44, 45). In essence, these papers defined what became, following the work of Michael Green and John Schwarz, superstring theory, and the projection has become known as the Gliozzi-Scherk-Olive, or GSO, projection, now one of the cornerstones of string theory. 

By 1976 David had come to accept that there was no real prospect of a permanent appointment at CERN and also that no effort was being made in Cambridge to find a tenured post there to which he could return, positions which may seem very difficult to understand given the significance his work was to acquire. David was offered a permanent post at the Niels Bohr Institute in Copenhagen, and he made a number of visits there, but he eventually concluded, for family reasons, that he would prefer to go back to the UK. Then, in 1977, he was offered a lectureship at Imperial College, London, and at the end of September, David and his family left Geneva to return to England. 

When David later reflected on his time in the Theory Division at CERN, he viewed it as a particularly happy and fruitful period, despite the shadow thrown towards the end by the temporary eclipse of string theory, and the implications that had for his job prospects. It was an exciting time to be there, with inspiring colleagues and a very pleasant atmosphere that was highly conducive to research. He came to regard his last academic year there, 1976-77, with GNO monopoles, the GSO projection, and Montonen-Olive duality, as the high point of his research career.

\bigskip
\bigskip
\bigskip
\centerline{\sc Imperial College}
\nopagebreak
\bigskip
While his family returned to their house in Barton, near Cambridge, which they had let while at CERN, David at first lodged in a house near Kew Station in south west London, convenient for travel to Imperial. He chose Kew because his father had lived there at one stage and nearby Richmond had family associations. Soon, however, he found a house to buy in Kew, into which he moved in mid March 1978. He spent the weekdays in London, returning to Cambridge at weekends, a pattern he was to follow for much of the next fifteen years, except when on leave. This gave him ample opportunity to follow his musical interests in the evenings, often visiting Covent Garden for the opera and ballet, and the Royal Festival Hall for concerts.

Soon after his arrival at Imperial College, David heard that his paper with Claus Montonen (47) had been accepted for publication. Although this paper was to become famous, David had been unhappy that they had not been able to find tighter arguments, and had worried that it might not be accepted by the journal. He wrote to his collaborator, ``I am glad Physics Letters accepted our paper without any embarrassment, but further papers must be more solid." Later it seemed to Montonen that this raised the bar so high that, in spite of many future discussions, no further papers were ever written.

One day in April 1978, David answered a knock on his office door to find Edward Witten, then a Junior Fellow at Harvard, who was visiting Oxford at the time at Michael Atiyah's invitation. Atiyah had told Witten that he thought that there might be something deep in David's work on monopoles, and advised him to seek David out at Imperial. Witten, who had not come across these papers before, was inclined to be skeptical of very speculative conjectures like that of Montonen-Olive. The key observation supporting the duality conjecture was that the same formula gave both the masses of the elementary electric charged particles and those of the magnetic monopoles, which are extended objects, and Witten could not see how this could survive the renormalization procedure necessary to define the quantum field theory. 

It occurred to them in discussion that there was some hope that supersymmetry might provide a rescue. By the end of the day of Witten's visit, they had understood that the supersymmetry algebra in a supersymmetric quantum field containing monopole solutions is modified by the presence of central charges. They were then able to show that, in the context of certain supersymmetric gauge theories, the supersymmetry algebra actually implies the Montonen-Olive mass formula, so that the prime result motivating the duality conjectures necessarily holds in suitable theories with supersymmetry. 

The joint paper (50) that they wrote was to become another of David's most influential contributions. Witten much later remarked that although he was very pleased with the result, with hindsight, he could see that he drew the wrong conclusion from it. In effect using Occam's razor, he took the fact that supersymmetry implies the Montonen-Olive mass formula to remove the need for any deeper explanation, such as duality. It was not until the work of Ashoke Sen (1994), showing the existence of certain two monopole states in N = 4 supersymmetric gauge theories, predicted by Montonen-Olive duality, that Witten became convinced of the depth and importance of the conjecture and, in large part through his influence, it made a seminal contribution to the reconceptualizing of string theory in the mid 1990s, in what has become known as ``the second superstring revolution''. 

During his time at Imperial, David would usually go back at weekends to the house he and Jenny had kept in Barton, near Cambridge, and often, on Saturday mornings, he would meet with Peter Goddard, by then a lecturer in DAMTP, to discuss physics in Goddard's office. These discussions provided the basis for a continuing collaboration that lasted from the late 1970s to the early 1990s. Soon after David moved to Imperial, they completed a review (49) on {\it Magnetic monopoles in gauge field theories}, which provided a systematic account of the ideas David had been centrally involved in developing in the previous four years. For some years thereafter, David was much in demand as a review speaker on this subject at conferences and summer schools.

In his first years at Imperial, David's research mainly focused on studying magnetic monopole solutions to spontaneously broken gauge theories in greater detail, in particular the conditions 
\begin{wrapfigure}{r}{105mm}
\vskip-6pt
\centering
  \includegraphics[width=100mm]{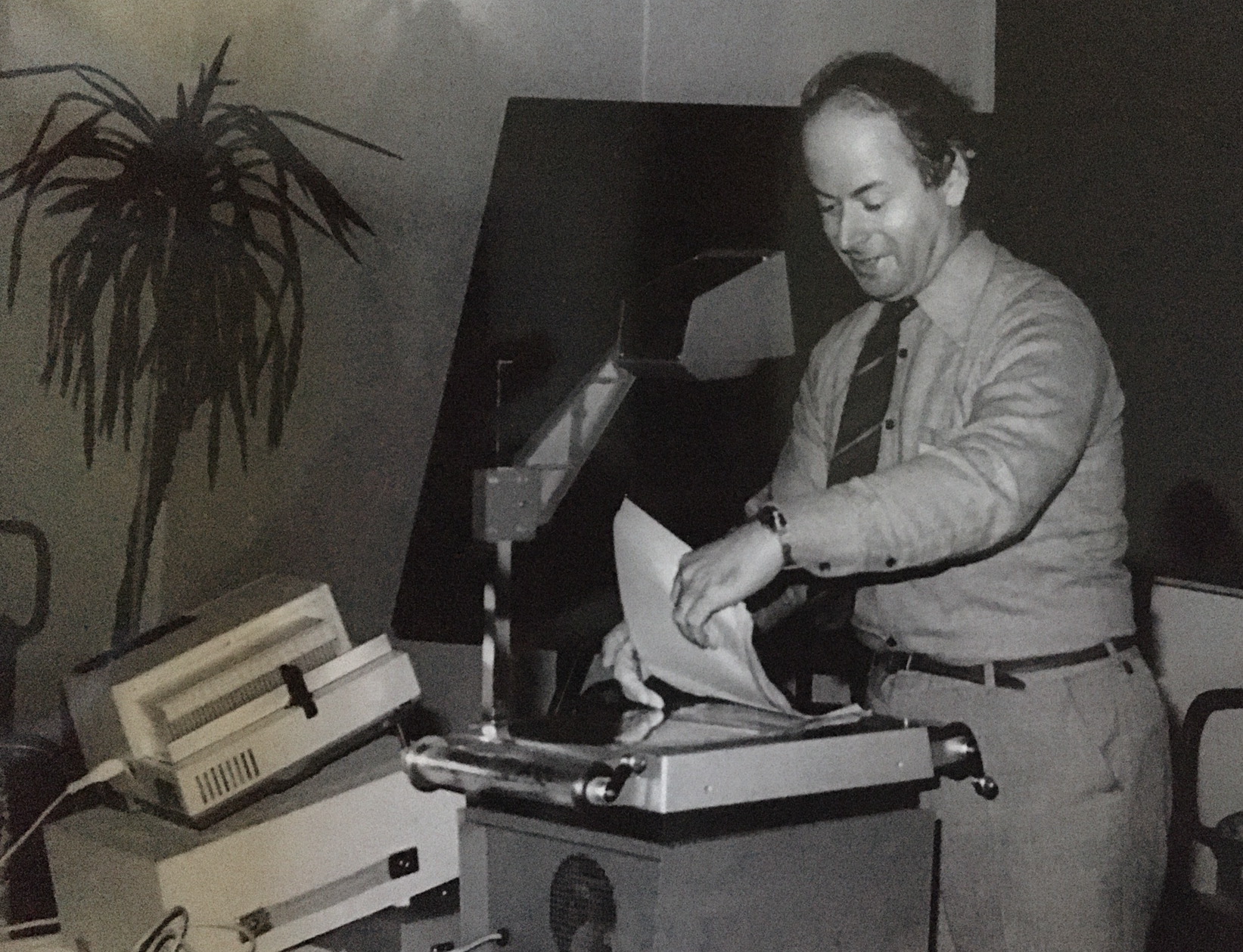}

\vskip2pt  
\scriptsize David Olive in Bechyn\v e, Czechoslovakia, in June 1981. \\
Photograph
kindly provided by the Olive family.
\vskip-18pt
\end{wrapfigure}
on their charges for their stability, finding further circumstantial evidence for the Montonen-Olive duality conjectures (60, 61). In general,  the nonlinear equations describing magnetic monopoles cannot be solved exactly in closed form, but it had been found that if attention is restricted to spherically symmetric solutions in an appropriate limit, the equations become integrable. Leznov and Saveliev (1979) had observed that where the gauge symmetry group is SU($N$), the equations are just those associated with the (finite) lattice of particles in a line  interacting through suitable nonlinear springs introduced fifteen years earlier by Morikazu Toda (1967). For a general (semisimple) gauge symmetry group, $G$, the equations can be characterized in terms of the Dynkin diagram of $G$, which encodes the group's structure (65). 

The way the study of spherically symmetric monopoles brought together solutions to gauge theories, the theory of Lie algebras and integrable systems struck David as deep and important. It was the initial motivation for his continuing interest in the Toda equations and their algebraic properties, a subject he returned to repeatedly over the next dozen years. At Imperial, David could easily attract and take on graduate students, in a way that had not been possible at CERN, and his work on Toda theories and other research was done in collaboration with a succession of current and former students, including  Regina Arcuri, Luiz Ferreira, Andreas Fring, Frank Gomes, Marco Kneipp,  Peter Johnson and Jonathan Underwood, and most notably Neil Turok.  Turok, who was later to hold professorships in both Princeton and Cambridge, before becoming the Director of the Perimeter Institute in Waterloo, Ontario, started as David's research student in 1980.  Over the following thirteen years he wrote a series of influential papers with David on Toda field theories ({\it e.g.} 70, 75, 102).

\bigskip
\bigskip
\bigskip

\centerline{\sc Leave in Charlottesville}
\nopagebreak
\bigskip
When David went on leave to the University of Virginia in Charlottesville in 1982-83, he took Neil Turok with him, carrying on their work on Toda theories. In Charlottesville, David gave a course of graduate lectures on the theory of Lie algebras, the mathematics that had underlain his seminal work on magnetic monopoles, but otherwise he used the freedom of his visiting appointment to develop the ideas that he had been evolving over some years in collaboration with Peter Goddard, who joined him there for some months at the beginning of 1983. Through discussions with Graeme Segal in Oxford and others, they had become aware that evidence of deep connections had begun to emerge between string theory and the theory of Kac-Moody algebras, a mathematical theory which had been initiated in 1967 by Victor Kac (1967) and Robert Moody (1967), coincidentally just about when the seeds of string theory were being sown by Veneziano's famous paper. 

Around 1980, mathematicians realized that, in constructing representations of Kac-Moody algebras, they had rediscovered the vertex operators that physicists had been using to describe the interactions of strings. The connection to the formalism of dual models was spelled out by Igor Frenkel  and Victor Kac (Frenkel \& Kac 1980) and by Segal (1981). The algebraic properties of these vertex operators had already featured prominently in work of both Goddard and Olive, and had been a central tool in understanding the structure of dual models and their detailed interpretation as string theory. Together in Charlottesville, and stimulated by what they had learned in the context of magnetic monopoles about the weight and root lattices of Lie algebras, Goddard and Olive studied how the vertex operator construction could be used to associate a Lie algebra to each integral lattice (i.e., a lattice such that the scalar product of any two lattice points is an integer). The nature of the Lie algebra so defined depended on whether the lattice is Euclidean or has some other signature. Although it seemed clear that these results should have some role to play in string theory, by producing symmetries of the spectrum, for example for strings moving in a space in which some of the dimensions have been compactified to form a torus formed by assuming periodicity under displacements corresponding to the lattice. However, at first sight, there seemed to be obstacles to using this to incorporate symmetry into string theory in a realistic way. 

Goddard and Olive paid particular attention to even self-dual lattices, in part because of their connection to modular invariance, which had already proved important in string theory. They noted that, in the Euclidean case, these lattices only existed in dimensions that were multiples of 8. In dimension 8, there was only one, the root lattice of the group $\Er_8$, while in dimension 16, there were two, the root lattice of $\Er_8 \times \Er_8$ and a sublattice of the weights of the Lie algebra of SO$(32)$. This observation proved prescient because just over a year later,  in the autumn of 1984, the 16-dimensional even self-dual Euclidean lattices, and the associated vertex operator construction of Kac-Moody algebras, were to play a key role in the developments that led to the dramatic revival of interest in string theory.

When Igor Frenkel visited Charlottesville in March 1983, Goddard and Olive were able to discuss their results with him and they discovered a strong overlap with results he had recently obtained. Frenkel arranged for them to be invited to speak at the conference on Vertex Operators in Mathematics and Physics that was held the following November in Berkeley at the recently established Mathematical Sciences Research Institute. Afterwards they circulated the paper, {\it Algebras, Lattices and Strings} (72),  which they had prepared for the proceedings and which was then influential in introducing many physicists to Kac-Moody algebras and their construction in terms of the vertex operators of string theory.

\bigskip
\bigskip
\bigskip
\centerline{\sc Coset Construction and Conformal Field Theory}
\bigskip
At the Berkeley conference, they also learned from the talk of Daniel Friedan about his work with Qiu and Shenker, and about then unpublished work of Belavin, Polyakov and Zamolodchikov, on conformal field theory and the representations of the Virasoro algebra. The Virasoro algebra is the Lie algebra of the conformal symmetry group and plays a central role both in string theory and in conformal field theory; indeed, in a sense, conformal field theory describes the structure of string theory. Friedan, Qiu and Shenker (FQS) had established necessary conditions for a representation of the Virasoro algebra to be unitary, showing that there was a continuum and an infinite discrete series (Friedan {\it et al.}~1984). But, apart from the two representations corresponding to a free fermion and to a free boson, a construction, or even a proof of existence, of the rest of the discrete series was wanting.

In the summer of 1983, David had returned to Imperial after he and his family had spent some weeks in the summer at the Aspen Center for Physics,  and he had resumed his Saturday meetings with Peter Goddard. The following spring their discussions focused on a recent paper of Edward Witten (1984), {\it Non-abelian bosonization in two dimensions}, which demonstrated an equivalence between certain boson theories, associated with Lie groups, now called Wess-Zumino-Witten (WZW) theories, and certain free fermion theories, by exploiting the equivalence of representations of isomorphic Kac-Moody algebras contained in the two theories. Looking for general conditions under which Witten's equivalence might hold, they sought to demonstrate that the energy-momentum tensors of the boson and fermion  theories were the same. In such two-dimensional theories, the moments of energy-momentum tensor generate the conformal symmetry of the theory, and they provide a representation of the Virasoro algebra. The aim of Goddard and Olive was to determine when the representations of the Virasoro algebra in the two theories were equivalent because this would be a necessary condition for the complete equivalence of the theories. 

Goddard and Olive spent some weeks working together at the Aspen Center for Physics in the summer of 1984, and found many instances where the two energy-momentum tensors differed. However, they realized that the difference between them was often very interesting. The boson theory could always be imbedded within the fermion theory and the question of equivalence was whether it was, in effect, all of that theory: subtracting the boson energy-momentum tensor from the fermion one gave zero if the theories were equivalent but otherwise the difference was itself an energy momentum tensor, and, in a sense described, what needed to be added to the boson theory to yield the full free fermion theory. The difference provided a representation of the Virasoro algebra, one which commuted with the boson representation of the Virasoro algebra, and such that their sum equaled the fermion representation. Further, explicit calculation showed that, in many instances, these `difference' Virasoro representations provided missing representations from the discrete series of FQS, thus proving their existence and giving an explicit construction in these cases. 

Goddard and Olive (74) did not manage to construct the whole of the infinite discrete series while in Aspen, but on their return to England they explained what they had done to Goddard's research student, Adrian Kent, and together they generalized the construction by replacing the fermion theory and considering instead a WZW theory associated with a Lie group, $G$, and the theory contained within it associated with a 
subgroup, $H$, of $G$. Any such pair, $H \subset G$, defines a Virasoro representation given by the difference of the Virasoro representations associated with the $G$ 
\begin{wrapfigure}{r}{105mm}
\vskip-12pt
\centering
   \includegraphics[width=100mm]{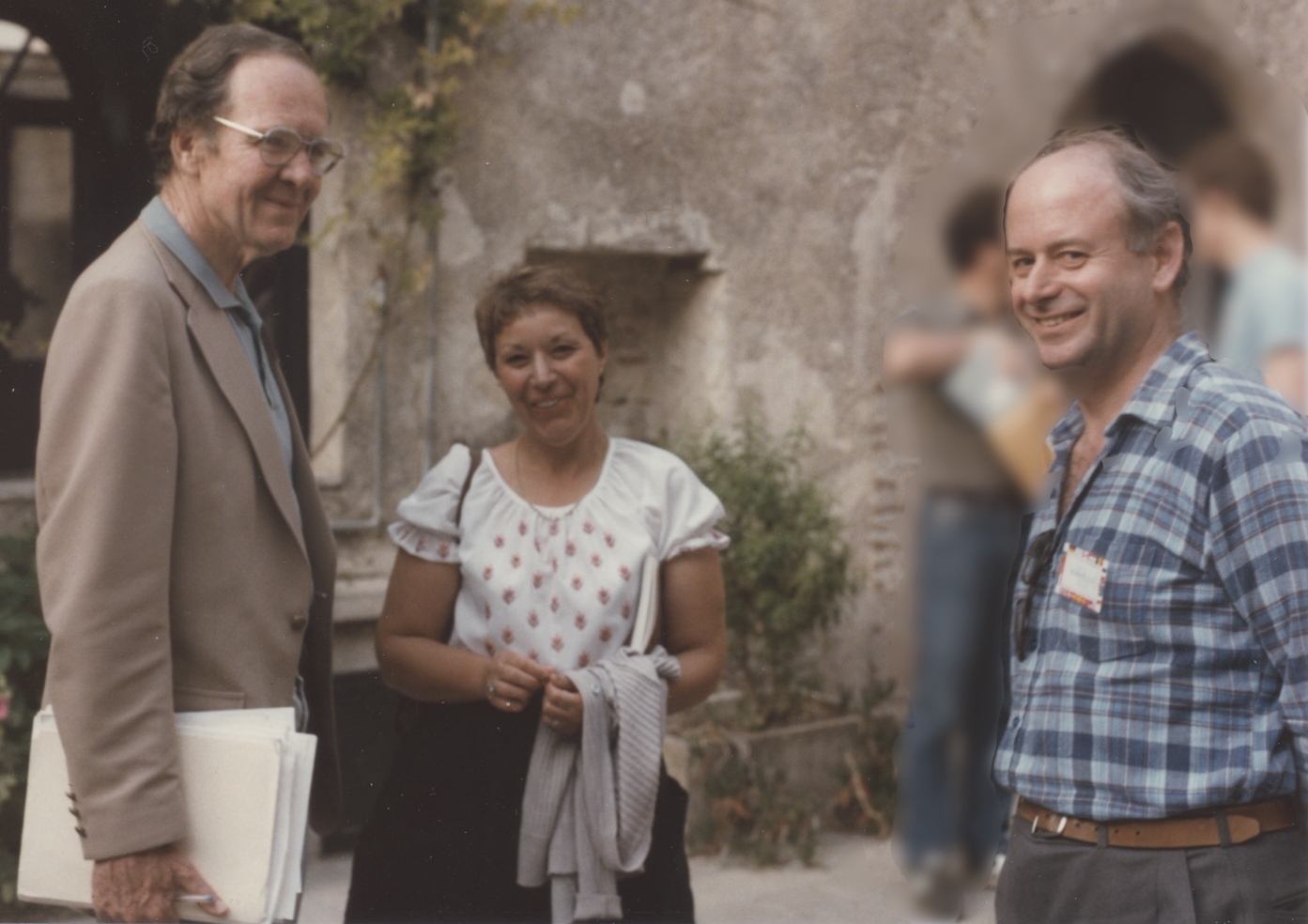}

\vskip2pt  
\scriptsize  David Olive with Arthur and Ludmilla Wightman at a summer school in Erice, Sicily, in August, 1985. Photograph
kindly provided by Arthur Jaffe.
\end{wrapfigure}
and $H$ theories. Goddard, Kent and Olive (GKO) associated this representation with the coset $G/H$ and it has become known as the coset or GKO construction (76, 81\aa). They were able to show that all the representations of the FQS discrete series could be obtained in this way, thus completing the classification of unitary representations of the Virasoro algebra. The WZW theory associated with a Lie group G is a conformal field theory and, for $H \subset G$, the coset construction associates a conformal field theory associated with $G/H$ with the corresponding energy-momentum tensor. This coset construction has remained one of the main ways of constructing and characterizing conformal field theories. 

The following year, 1985, in collaboration with Werner Nahm, Goddard and Olive succeeded in solving the original problem from which they had been fruitfully diverted by the coset construction, by showing that the condition for the fermion theory in Witten's non-Abelian bosonization to be equivalent to a WZW theory was that the fermions should transform according to a representation of $G$ that could be used to extend it to form a larger group, $G'$, in such a way that the pair define what is called a symmetric space (78).

\bigskip
\bigskip
\bigskip
\centerline{\sc Revival of String Theory}
\bigskip

Also at Aspen in the summer of 1984 were Michael Green and John Schwarz, who were studying anomaly cancellations in supersymmetric gauge theories coupled to gravity in the hope of establishing that, under suitable conditions, string theory might satisfy this consistency requirement. 
They first found that the cancellation occurred for the gauge group with the Lie algebra of SO$(32)$, provided that its weights lay on the lattice which is even and self-dual, and then they realized that it also holds for $\Er_8 \times \Er_8$ (Green \& Schwarz 1984). Aware that these two groups had been singled out by Goddard and Olive a year before in their paper, {\it Algebras, Lattices and Strings}, they sought to construct a string theory with  $\Er_8 \times \Er_8$ symmetry, unknown at the time, based on ideas from that paper on the incorporation of symmetry into string theory using Kac-Moody algebras. But further ideas were necessary, about treating left and right moving waves on closed strings very differently, which came with the construction by Gross, Harvey, Martinec and Rohm (1985) of what they called the heterotic string. 

It was these developments, initiated by the work of Green and Schwarz, and promulgated through the unique influence of Edward Witten, that gave rise to the renaissance of interest in string theory beginning in 1984. In them, Kac-Moody algebras, and two-dimensional conformal field theory more generally, played a key role. As David continued working with Peter Goddard, Werner Nahm, Adam Schwimmer and others on conformal field theory and related infinite-dimensional algebras, the mushrooming interest in string theory meant that he was much in demand as a lecturer at conferences and summer schools. This prompted him to write a long pedagogical review, {\it Kac-Moody and Virasoro Algebras in Relation to Quantum Physics} (82\aa), in collaboration with Peter Goddard, which was widely read and has remained a standard reference.  Alongside his work on conformal field theory, David continued to make contributions to Toda field theory with Turok and many of his other students. With a postdoctoral fellow, Michael Freeman, he also returned to the basic calculation of the one loop contribution in bosonic string theory, using the BRS formalism to simplify his earlier work with Lars Brink (87\aa, 88\aa).

From the mid 1980s, David began to receive long overdue formal recognition of his achievements. He had been made a Reader at Imperial in 1980 at the age of 43, and four years later he was promoted to a Professorship of Theoretical Physics. In 1987, he was elected a Fellow of the Royal Society in recognition of his contributions to $S$-matrix theory, dual models, the classification of magnetic monopoles in gauge field theories and on the concept of electric-magnetic duality.
 
David spent the academic year 1987-88 as a Member of the Institute for Advanced Study in Princeton. When he returned to Imperial, he became head of the theoretical physics group, but the administrative duties of this post did not really suit his temperament.  He discharged his responsibilities perhaps too meticulously,  even down to counting out rations of photocopying paper for students when  there was a funding crisis. The very qualities of mind, the almost obsessive need to resolve possibly tell-tale discrepancies in understanding, that led to David's remarkably original and prescient research, nearly drove David and those around him to distraction when he tried to reconcile, down to the last penny, the different financial systems Imperial deployed at the group, department and college levels. When a new administrator with some understanding of accounts arrived to support the group, it was not only David who was enormously relieved.

\bigskip
\bigskip
\bigskip

\centerline{\sc Swansea and Later Years}

\bigskip
In 1991, when the University College of Wales Swansea was seeking to fill a chair in physics, Aubrey Truman, a mathematical physicist and the Dean of Science, argued in favour of appointing a theoretical physicist, particularly in view of the fertile cross connections with mathematics that had developed in the previous two decades in particular. Truman had got to know David Olive when David had served on the scientific advisory committee for the International Congress of Mathematical Physics that had taken place in Swansea in 1988. They had both been research students of JC Taylor, Olive in Cambridge and Truman later in Oxford, and, although they had not met, Truman had long admired Olive's contributions to $S$-matrix theory as well as his later work. He approached David about the vacant chair and was delighted when he expressed an interest in moving from Imperial to Swansea. 

It turned out that David's colleague at Imperial, Ian Halliday, was also attracted by the idea of moving to Swansea. At the time the future of the Swansea Physics Department had looked quite uncertain, and the Vice Chancellor, Brian Clarkson, became interested in the idea of recruiting Olive and Halliday as the nucleus of a theoretical particle physics group that might substantially raise the international standing of the Physics Department. The plan was formed of appointing David as a Research Professor in the Department of Mathematics and Halliday in Physics, with the group they were to lead linking the two departments. However, this initial arrangement did not last, because there were tensions with some other members of the Department of Mathematics. David was not a political animal and he just got very agitated when he thought others were not behaving straightforwardly. As a result, the idea of linking the two departments was abandoned and the whole group  moved into Physics. 

David demonstrated his characteristic carefulness in the negotiations with the University over the group's move. Here his meticulousness proved an asset because he ensured that every last promised provision was 
written out in exquisite detail in letters signed by the Vice Chancellor, 
which proved extremely useful to the group in ensuring that commitments were kept. David was very engaged in the whole development of the new group and his presence was a decisive factor in the recruitment of a number of outstanding young lecturers, Nick Dorey, Tim Hollowood, Warren Perkins, and Graham Shore, soon followed by others. The new 
group brought highly valuable recognition to the physics department at a time when pure science was being cut in universities like Swansea; the ranking of the Physics Department improved dramatically. The group 
Halliday and Olive created, now with a strength of about twelve faculty members,  has continued to thrive  over the last twenty-five years  as one of the leading theoretical particle physics groups in the UK.

David had spent the second half of 1992 participating in one of the first programmes at the Newton Institute in Cambridge. When the Olives moved to Swansea at the beginning of 1993, they retained 
their home in Barton, near Cambridge. Although Jenny may have made the move somewhat 
\begin{wrapfigure}{l}{105mm}
\vskip-12pt
\centering
   \includegraphics[width=100mm]{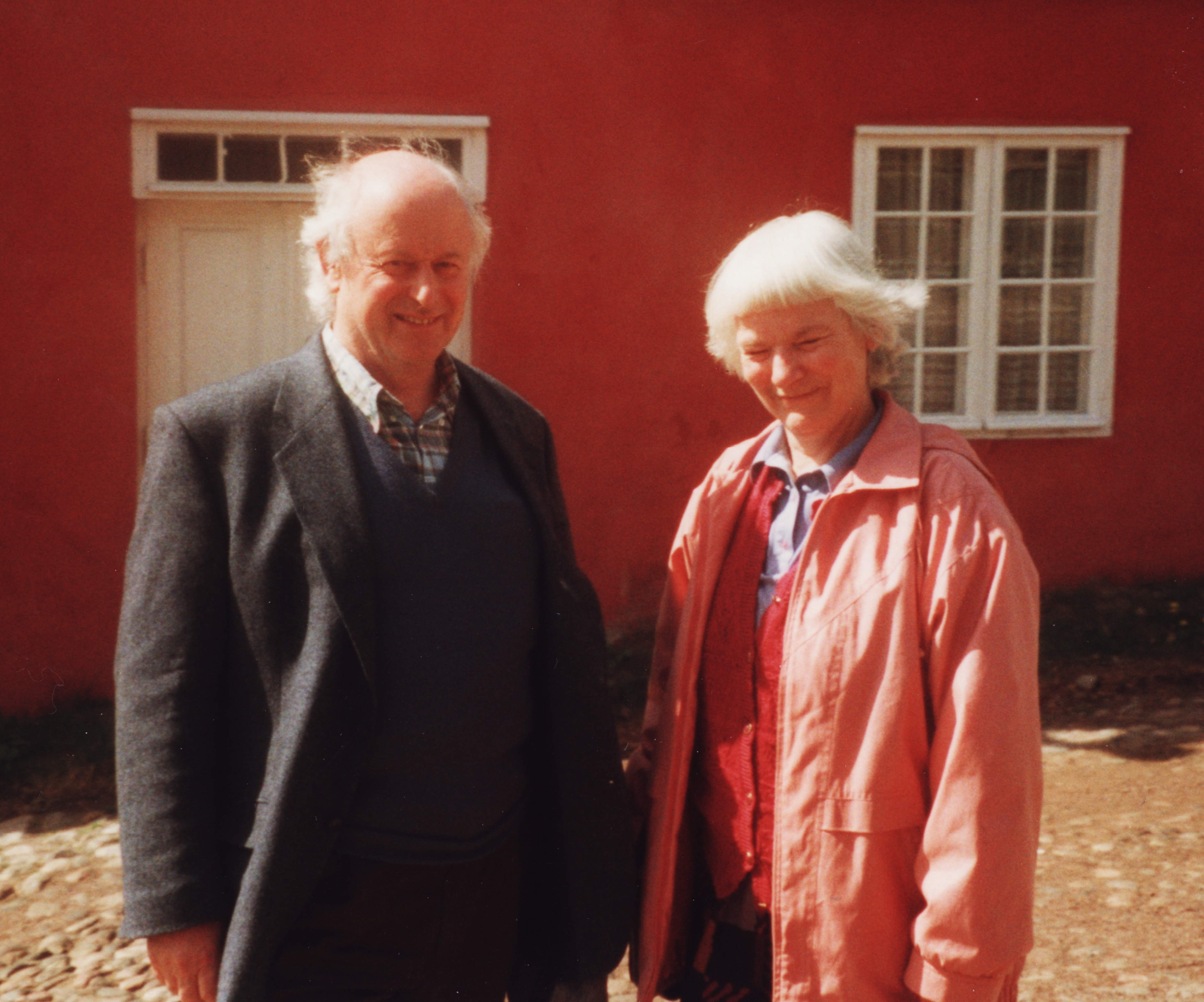}
\vskip2pt  
\scriptsize David and Jenny Olive in Finland in May 1999. Photograph
kindly provided by Claus Montonen.
\vskip-12pt
 \end{wrapfigure}
reluctantly, she continued to support David in his work, even helping with the production of diagrams for his public lectures. She also helped new physics students in the university who were having difficulties with basic mathematics. 
She shared the techniques she had developed for this with a wider audience in her book, {\it Maths: A Student's Survival Guide. A Self-Help Workbook for Science and Engineering Students}, published by Cambridge University Press.

David had had four Brazilian research students, Regina Arcuri, Luis Ferreira, Frank Gomes and Marco Kneipp. In these years he made a number of visits to Brazil, with Jenny sometimes accompanying him, to 
participate in workshops, research schools, {\it etc.}, contributing to furthering the development of theoretical physics in the country. He developed a strong interest in Brazilian culture, particularly the food and the 
music. He loved to have lunch in the restaurants close to the Paulista Avenue in S\~ao Paulo, where one pays by the weight of the food taken from a generous buffet. Characteristically,  David developed models for how to get best value for money. But it was Brazilian music that most engaged him, from Villa-Lobos to much less known composers, such as Lorenzo Fernandez. He became extremely interested in a recording of Fernandez's music by the Amazonia Quartet, but it seemed that it was sold out in S\~ao Paulo and had been discontinued. However, he tracked it down to a small shop in a distant part of the city  and so acquired a rare CD, of which he was very proud.

Further public recognition of David's achievements came with the award in 1997 of the Dirac Prize and Medal of the International Centre for Theoretical Physics in Trieste, shared with Peter Goddard, ``in recognition of their far-sighted and highly influential contributions to theoretical physics''. The announcement further cited their contribution of ``many crucial insights that shaped our emerging understanding of string theory and have also had a far-reaching impact on our understanding of 4-dimensional field theory'', and went on to explain that ``Olive's work on spacetime supersymmetry of the spinning string theory (with F. Gliozzi and J. Scherk) made possible the whole idea of superstrings, which we now understand as the most natural framework for supersymmetry and string theory. Goddard and Olive introduced key ideas about the use of current algebra in string theory which were very important in the subsequent discovery of attractive ways to incorporate space-time gauge symmetry in string theory, thus making it possible for string theory to incorporate the standard model of particle physics. These discoveries, made in the years 1973-83, were among the most crucial steps in making possible the `superstring revolution' of 1984-5. The `second superstring revolution' of the last few years has been equally dependent on pioneering insights about magnetic monopoles made in 1977 by Goddard, Olive, and J. Nuyts, and further extended by Olive and C. Montonen. Their ideas concerning a dual interpretation of magnetic charge, and then about electric-magnetic duality in non-abelian gauge theory, were way ahead of their time and have proved to have a far-reaching importance, which we are only now beginning to understand, in governing the dynamics of four-dimensional field theory and of superstring theory.'' 

The `second superstring revolution', to which the Dirac Medal citation referred, was  initiated by the work of Nathan Seiberg and Edward Witten (1984a,b). Central to it was the generalization of Montonen-Olive 
duality to string theory, in the form of S-duality, and its extension to a modular group of dualities by the work of Ashoke Sen (1994). After twenty years, it had at last been  realized  widely how visionary David's work 
on monopoles in the mid 1970s had been. These developments rekindled David's own interest in electromagnetic duality. With Pierre Van Baal and Peter West, David organized a programme at the Newton Institute on Non-Perturbative Aspects of Quantum Field Theory, centred on ideas of duality and supersymmetry, in the first six months of 1997 (116\aa). His deep and beautifully clear expositions of electromagnetic duality and its generalizations became in demand at conferences and graduate schools. With Marco Alvarez, who had joined him as a postdoctoral fellow in Swansea, in some of his last work, David characteristically sought a deeper understanding of the quantization of magnetic flux, by considering gauge theories on smooth four-dimensional manifolds of arbitrary topology. They found that the quantization of fluxes in these theories provided a physical interpretation of some mathematical concepts, such as the Stiefel-Whitney classes (119\aa, 122\aa, 130\aa).

David received national recognition at the beginning of 2002 with his appointment as a Commander of the Order of the British Empire, for services to theoretical physics, in the UK New Year Honours, and, in 2007, he was made a Foreign Member of the Royal Society of Arts and Sciences in Gothenburg, Sweden. 
In many ways, David enjoyed his time in Swansea more than his years at Imperial, which were made somewhat fraught by commuting on crowded trains and having to lecture to very large audiences of students. With his colleague Colin Evans, David  began to compile a history of the Swansea Physics Department and he became an expert on the history of Welsh science, particularly in relation to the history of cosmic rays, operations research, radar, the atomic bomb, and the internet. He was a Founding Fellow of the Learned Society of Wales in 2010, and a strong supporter of it. 

In the last six years of his life David suffered from increasingly poor health as a result of heart disease. He and Jenny first became aware of a problem when David became exceedingly tired when they were walking in the Lauterbrunnen Valley. Later, his golfing partner, a physician, advised him to consult his doctor because of the difficulties he was experiencing while they were playing. His deteriorating heart condition led eventually to kidney failure, but the weakness of his heart meant that dialysis was not possible. Through his last years, David remained mentally engaged, and, in spite of his difficulty in walking any distance, he continued to travel for special occasions and conferences up to a few months before his death. 

In late October 2012, David's health suddenly deteriorated  further and he and Jenny realized that he was gravely ill. One kidney had failed and efforts to keep the other kidney working were proving ineffective. His doctors concluded that nothing more could be done. David, then in hospital, phoned Jenny and told her in a rather matter of fact way that he was finally being allowed to come home for palliative care, because no further treatment would be effective. As was then expected, he only had a week to live, during which he showed grace and courage as he received many greatly appreciated telephone calls and emails  from physicists and mathematicians around the world, who had heard of his condition.

David died on 7 November in the house in King's Grove, Barton, that he and Jenny had owned since 1965. At his funeral service twelve days later, the village's 14th century parish church, St Peter's, was full with David's friends, colleagues and relatives. David did not have a religious belief, but the kindly vicar told the congregation that we could be sure that David was now in a place where there were no more physics problems to worry about, not aware that that would have been no paradise at all for David. Following his wishes, he was buried in the village churchyard. 

\bigskip
\bigskip
\bigskip
\centerline{\sc Interests, Personality and Influence}

\bigskip

David had a wide range of interests, but his life-long love of music was exceptional. From his youth, he built up an outstanding collection of long playing records, which were gradually replaced from the mid 1980s onwards by an even more extensive collection of over four thousand CDs, housed in nine cabinets, which he had constructed personally with great care and skill. He carried over from his LP collection his practice of meticulously making a note on the sleeve of each time he played a recording, in part to assess wear. 

He had an encyclopedic knowledge of music, and of his collection of recordings in particular, which often astonished his friends and colleagues.  Adam Schwimmer, who grew up in the little-known town of Grosswardein-Oradea in Transylvania, recalled a characteristic incident illustrating both the depth of David's musical knowledge and his particular sense of humour. One evening David produced from one of his CD cabinets a recording of symphonies by Michael Haydn, Joseph's much less famous younger brother, played by the Oradea Philarmonic in  what Schwimmer described as an awful performance. David knew that Michael Haydn had been the court composer of the Bishop of Grosswardein, Schwimmer's home town, and he had in his collection the only available recording of his works.

David's notes on and knowledge of the concerts and opera performances he had attended over the years paralleled, even rivaled, the notes he made setting out his evolving understanding of his investigations in theoretical physics.  Filed in numerous loose leaf folders, and annotated with underlinings in various colours and styles, which David did not even attempt to explain to others, these notes were written and rewritten as he sought an every deeper and clearer understanding of the topic at hand, whether this was an established branch of mathematics or a new physical theory. In either case, he would strive to gain his own understanding, one that met his own exacting standards of concision and clarity, often producing novel insights into existing theories.  

David's frequent comment was ``I think we can understand this better'', but, even when sceptical of an idea, he would not simply dismiss it but rather look for ways in which it might make sense. He was always seeking mathematically deep, elegant equations that encapsulated physical ideas. It was quite evident that his approach emulated naturally that of his great intellectual hero, Paul Dirac, whose lectures he had attended as a student at Cambridge. 

The many files of handwritten notes that he amassed seemed an end in themselves for him, a 
\begin{wrapfigure}{r}{105mm}
\vskip-12pt
\centering
  \includegraphics[width=100mm]{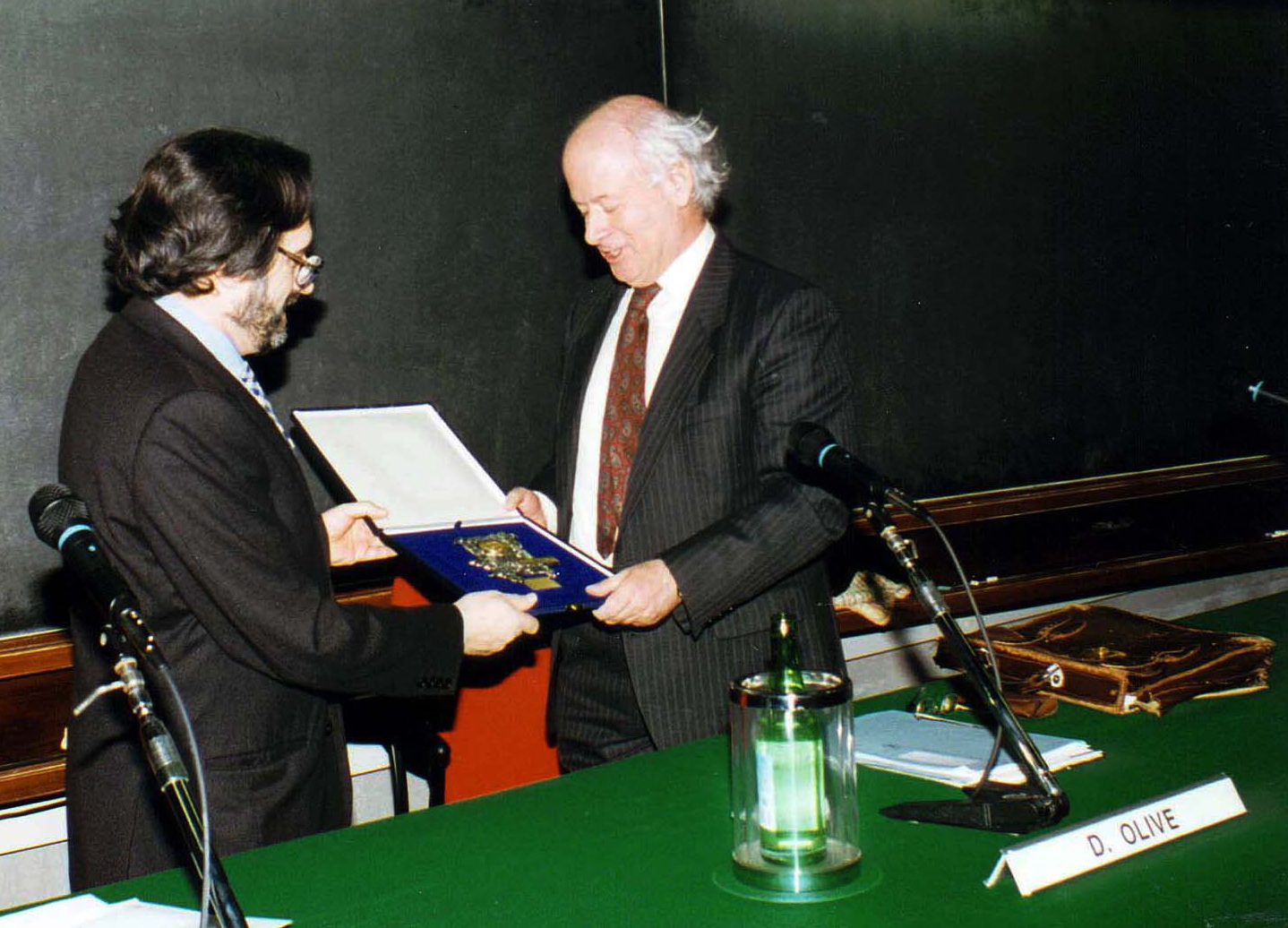}

\vskip2pt  
\scriptsize David Olive accepting the Dirac Medal from Miguel Virasoro, Director of 

ICTP,  in Trieste on 26 March 1998. The briefcase that David had used   

from his schooldays, for nearly fifty  years, can be seen at the right.
\end{wrapfigure}
record and reference for what he had understood. They would form the basis for his research publications, his lecture notes and his reviews, and their honing  underpinned David's exceptionally clear expository style,
but he would be very reluctant to publish a result before he was convinced that the argument was sufficiently clear, 
sometimes to the frustration of his colleagues. He was justifiably proud of the seminal contributions that he had made to theoretical physics, but he never had a desire to rush into print to seek priority on an idea that had not yet been elucidated to his own personal satisfaction.  
Although he was determined to understand everything himself {\it de novo}, he readily shared the results of his endeavours with others. 

The openness and generosity with which David would share his new ideas and insights with his collaborators, colleagues and students was in amusing contrast to 
the minute care with which he would dissect the bill for a meal shared with a colleague, for example in one of the Indian restaurants he loved in London, evoking the cultural stereotype of a Scotsman. The leather briefcase, acquired as a schoolboy at the Royal High School, Edinburgh, was retained until nearly the end of his career. Progressively disintegrating, but kept in service by David's own highly imaginative repairs, performed with characteristic economy, e.g. by use of bits from an old Fairy Liquid bottle to mend the handle, it provided an outward symbol of David's ingenuity and endearing frugality and it was only abandoned with the greatest reluctance in his later years.

For some years, his entry in {\it Who's Who} seemed to contain a perhaps Freudian reference to his concern to conserve financial resources: his interests were listed there as  `music and gold', the latter a misprint occasioned by his at times almost illegible handwriting; eventually it was corrected to `golf', the other passion, together with music, that he carried 
throughout his life from his teenage years onwards. In Swansea, he was able to find the opportunities to play the game to an extent that he had not enjoyed since his youth in Scotland. He and Ian Halliday joined Pennard Golf Club, and for many years they played together every Saturday morning. Halliday found David's golf style to be the antithesis of his careful, though inspirational, approach to physics: he would occasionally hit  enormous drives but with very little control,  and on one memorable occasion his drive hit the ladies' tee marker thirty  
yards ahead and, following a trajectory reminiscent of Rutherford scattering,  the ball whistled back past Halliday's and Olive's ears to come to rest behind them, a drive of minus 150 yards. 

He was in many ways a quiet and reserved person but he enjoyed the company of his friends and colleagues and he loved the process of collaboration, as the papers containing his major contributions to physics make clear. These were ahead of their time and helped to shape the understanding of the structures of string theory and quantum field theory gained in the last forty years.
  His relentless and uncompromising search for rigour and clarity permanently influenced all those he taught or worked with.

\bigskip
\bigskip
\bigskip
\vfil\eject
\centerline{\large\sc II\quad Scientific Contributions}
\bigskip
\bigskip
\centerline{\sc $S$-matrix Theory}
\bigskip
\nobreak
{\it The Analytic S-Matrix} (12) begins with the rather arch sentence, ``One
of the most important discoveries in elementary particle physics has been that of
the complex plane.'' The  analyticity of two-particle scattering amplitudes in energy, treated as a complex variable, had been used for some decades to derive dispersion relations, expressing the amplitude as an integral of  its imaginary part, under suitable assumptions. However, just before David began research at the beginning of the 1960s, Stanley Mandelstam (1959) showed how the scattering amplitudes for any number of particles could be considered as analytic functions of the invariants formed from the momenta of the particles. He demonstrated that the amplitudes had singularities, poles and branch cuts, whose presence could be seen as following directly from the interrelation of analyticity and unitarity, and that the perturbation series in a quantum field theory could be reconstructed from these requirements. 

Building on Mandelstam's ideas, Geoffrey Chew formulated the bootstrap hypothesis, that is the proposition that the requirements of analyticity (reflecting causality) and unitarity (seemingly essential for the probabilistic interpretation of quantum theory), together with some asymptotic assumptions at high energy, determine the scattering amplitudes, {\it i.e.} the $S$-matrix, uniquely. For Chew's most zealous followers, the bootstrap hypothesis became almost an article of faith, and Chew was a charismatic evangelist for it himself. For others, including Mandelstam as well as Olive, it remained a hypothesis that merited exploration. 

David's first contributions in this area (2--4,6) were concerned relations between analytic continuations, discontinuities, and unitarity. At first, he worked within the context of quantum field theory, progressing in (6)  to discuss the possibility of derivations within an $S$-matrix theory, whose structure he had begun to explore in {\it An exploration of $S$-matrix theory} (5). In {\it The Analytic $S$ Matrix} (1966), published at the same time as (12), but without the hyphen, Chew gave his fullest account of his philosophy. He cited David's work (5), as the most ambitious attempt to provide an axiomatic treatment of the programme, summarizing his treatment and employing the `bubble' notation that David had developed and which became standard in the subject.    

David built on earlier work of Stapp (1962) and, particularly, that of Gunson (1965), which was available as a preprint in 1963. The main ingredient that had to be included as a starting point, in addition to analyticity and unitarity, was connectedness structure, taken to be a consequence of the assumed short-range nature of the fundamental forces. For the simplest case of two-to-two scattering, $a+b\rightarrow a+b$, with momenta $p_a,p_b$ in the initial state and $p'_a,p'_b$ in the final state, the statement of connectedness in `bubble' notation is

\vfil\eject
\begin{center}
\begin{tikzpicture}[scale=0.5]
\def \x {0}
\draw [thick] (\x,0) circle (1);
\draw [thick] (\x-2,0.5) --(\x-0.9,0.5);
\draw [thick] (\x-2,-0.5)--(\x-0.9,-0.5);
\draw [thick] (\x+0.9,0.5) --(\x+2,0.5);
\draw [thick] (\x+0.9,-0.5)--(\x+2,-0.5);
\node  at (0,0) {$S$};
\def \x {4}
\draw [thick] (\x-0.4,0.15) --(\x+0.4,0.15);
\draw [thick] (\x-0.4,-0.15) --(\x+0.4,-0.15);
\def \x {6}
\draw [thick] (\x,0.5) --(\x+2,0.5);
\draw [thick] (\x,-0.5)--(\x+2,-0.5);
\def \x {10}
\draw [thick] (\x-0.4,0) --(\x+0.4,0);
\draw [thick] (\x,-0.4) --(\x,0.4);
\def \x {14}
\draw [thick] (\x,0) circle (1);
\draw [thick] (\x-2,0.5) --(\x-0.9,0.5);
\draw [thick] (\x-2,-0.5)--(\x-0.9,-0.5);
\draw [thick] (\x+0.9,0.5) --(\x+2,0.5);
\draw [thick] (\x+0.9,-0.5)--(\x+2,-0.5);
\draw [thick] (\x-0.4,0) --(\x+0.4,0);
\draw [thick] (\x,-0.4) --(\x,0.4);
\node [left] at (-2,0.5) {$a$};
\node [left] at (-2,-0.5) {$b$};
\node [right] at (2,0.5) {$a$};
\node [right] at (2,-0.5) {$b$};
\node [left] at (6,0.5) {$a$};
\node [left] at (6,-0.5) {$b$};
\node [right] at (8,0.5) {$a$};
\node [right] at (8,-0.5) {$b$};
\node [left] at (12,0.5) {$a$};
\node [left] at (12,-0.5) {$b$};
\node [right] at (16,0.5) {$a$};
\node [right] at (16 ,-0.5) {$b$};
\end{tikzpicture}
\end{center}
\vspace{-14truemm}
\be\label{s1}\phantom{x}\ee

where the lefthand side denotes the $S$-matrix element $\langle p'_a,p_b'|S|p_a,p_b\rangle$ and, on the righthand side, the first term denotes  $\langle p'_a,p_b'|p_a,p_b\rangle$, essentially the product of delta functions between initial and final three-momenta for each of the two particles $a$ and $b$, and, up to normalization factors, the second term equals $i\delta_4(p_a+p_b-p_a'-p_b')A$. Here $A$ is the `connected' part of the scattering amplitude, a function of the invariants $s=(p_a+p_b)^2$ and $t=(p_a-p_a')^2$, which is assumed to be as analytic as possible, consistent with the requirements of unitarity, 

Unitarity for the $S$-matrix, $S$, takes the form $SS^\dagger=1$, which in `bubble' notation reads
\begin{center}
\begin{tikzpicture}[scale=0.5]
\def \x {0}
\draw [thick] (\x,0) circle (1);
\draw [thick] (\x-2,0.5) --(\x-0.9,0.5);
\draw [thick] (\x-2,-0.5)--(\x-0.9,-0.5);
\draw [thick] (\x+0.9,0.5) --(\x+2.1,0.5);
\draw [thick] (\x+0.9,-0.5)--(\x+2.1,-0.5);
\node  at (0,0) {$S$};
\def \x {3}
\draw [thick] (\x,0) circle (1);
\draw [thick] (\x+0.9,0.5) --(\x+2,0.5);
\draw [thick] (\x+0.9,-0.5)--(\x+2,-0.5);
\node  at (3.1,0.02) {$S^\dagger$};
\def \x {6.5}
\draw [thick] (\x-0.4,0.15) --(\x+0.4,0.15);
\draw [thick] (\x-0.4,-0.15) --(\x+0.4,-0.15);
\def \x {8}
\draw [thick] (\x,0.5) --(\x+2,0.5);
\draw [thick] (\x,-0.5)--(\x+2,-0.5);
\end{tikzpicture}
\end{center}
\vskip-43pt\be\label{s2}\phantom{x}\ee\vskip6pt
where the lines joining indicating integration over momenta and the particle labels have been omitted.
So, substituting for $S$ from \eqref{s1},
\begin{center}
\begin{tikzpicture}[scale=0.5]
\def \x {0}
\draw [thick] (\x,0) circle (1);
\draw [thick] (\x-2,0.5) --(\x-0.9,0.5);
\draw [thick] (\x-2,-0.5)--(\x-0.9,-0.5);
\draw [thick] (\x+0.9,0.5) --(\x+2,0.5);
\draw [thick] (\x+0.9,-0.5)--(\x+2,-0.5);
\draw [thick] (\x-0.4,0) --(\x+0.4,0);
\draw [thick] (\x,-0.4) --(\x,0.4);
\def \x {3}
\draw [thick] (\x-0.4,0) --(\x+0.4,0);
\def \x {6}
\draw [thick] (\x,0) circle (1);
\draw [thick] (\x-2,0.5) --(\x-0.9,0.5);
\draw [thick] (\x-2,-0.5)--(\x-0.9,-0.5);
\draw [thick] (\x+0.9,0.5) --(\x+2,0.5);
\draw [thick] (\x+0.9,-0.5)--(\x+2,-0.5);
\draw [thick] (\x-0.4,0) --(\x+0.4,0);
\def \x {9.5}
\draw [thick] (\x-0.4,0.15) --(\x+0.4,0.15);
\draw [thick] (\x-0.4,-0.15) --(\x+0.4,-0.15);
\def \x {13}
\draw [thick] (\x,0) circle (1);
\draw [thick] (\x-2,0.5) --(\x-0.9,0.5);
\draw [thick] (\x-2,-0.5)--(\x-0.9,-0.5);
\draw [thick] (\x+0.9,0.5) --(\x+2.1,0.5);
\draw [thick] (\x+0.9,-0.5)--(\x+2.1,-0.5);
\draw [thick] (\x-0.4,0) --(\x+0.4,0);
\draw [thick] (\x,-0.4) --(\x,0.4);
\def \x {16}
\draw [thick] (\x,0) circle (1);
\draw [thick] (\x+0.9,0.5) --(\x+2,0.5);
\draw [thick] (\x+0.9,-0.5)--(\x+2,-0.5);
\draw [thick] (\x-0.4,0) --(\x+0.4,0);
\end{tikzpicture}
\end{center}
\vskip-43pt\be\label{s3}\phantom{x}\ee\vskip6pt
where the second term on the lefthand side is the hermitian conjugate of the first. In (2), within the context of quantum field theory, David had argued that this hermitian conjugate, the `minus' amplitude, is the analytic continuation of the first term in [\ref{s3}], the `plus' amplitude, through the unphysical region below $s=(m_a+m_b)^2$, the two-particle normal threshold, more generally than had previously been established. [Here $m_a, m_b$ denote the masses of $a,b$, and so on.] In (5), he produced a more general argument for the validity of this property, known as hermitian analyticity, within the context of $S$-matrix theory. [See also (12), p. 223.] For $s$ real just below this threshold and for suitable real $t$, the `plus' and `minus' amplitudes
are equal and real, and so  [\ref{s3}] holds with the expression on the righthand side replaced by zero. 

Above the two-particle threshold,  [\ref{s3}] gives an expression for the discontinuity across a cut, with the `plus' and `minus' amplitudes being the boundary values of a single analytic function from different sides of the cut. If, at higher energy, it becomes possible to produce a particle, $c$,  {\it i.e.} $a+b\rightarrow a+b +c$, an extra term appears in the unitarity equation, which takes the form
\begin{center}
\begin{tikzpicture}[scale=0.5]
\def \x {0}
\draw [thick] (\x,0) circle (1);
\draw [thick] (\x-2,0.5) --(\x-0.9,0.5);
\draw [thick] (\x-2,-0.5)--(\x-0.9,-0.5);
\draw [thick] (\x+0.9,0.5) --(\x+2,0.5);
\draw [thick] (\x+0.9,-0.5)--(\x+2,-0.5);
\draw [thick] (\x-0.4,0) --(\x+0.4,0);
\draw [thick] (\x,-0.4) --(\x,0.4);
\def \x {3}
\draw [thick] (\x-0.4,0) --(\x+0.4,0);
\def \x {6}
\draw [thick] (\x,0) circle (1);
\draw [thick] (\x-2,0.5) --(\x-0.9,0.5);
\draw [thick] (\x-2,-0.5)--(\x-0.9,-0.5);
\draw [thick] (\x+0.9,0.5) --(\x+2,0.5);
\draw [thick] (\x+0.9,-0.5)--(\x+2,-0.5);
\draw [thick] (\x-0.4,0) --(\x+0.4,0);
\def \x {9.5}
\draw [thick] (\x-0.4,0.15) --(\x+0.4,0.15);
\draw [thick] (\x-0.4,-0.15) --(\x+0.4,-0.15);
\def \x {13}
\draw [thick] (\x,0) circle (1);
\draw [thick] (\x-2,0.5) --(\x-0.9,0.5);
\draw [thick] (\x-2,-0.5)--(\x-0.9,-0.5);
\draw [thick] (\x+0.9,0.5) --(\x+2.1,0.5);
\draw [thick] (\x+0.9,-0.5)--(\x+2.1,-0.5);
\draw [thick] (\x-0.4,0) --(\x+0.4,0);
\draw [thick] (\x,-0.4) --(\x,0.4);
\def \x {16}
\draw [thick] (\x,0) circle (1);
\draw [thick] (\x+0.9,0.5) --(\x+2,0.5);
\draw [thick] (\x+0.9,-0.5)--(\x+2,-0.5);
\draw [thick] (\x-0.4,0) --(\x+0.4,0);
\def \x {19}
\draw [thick] (\x-0.4,0) --(\x+0.4,0);
\draw [thick] (\x,-0.4) --(\x,0.4);
\def \x {22}
\draw [thick] (\x,0) circle (1);
\draw [thick] (\x-2,0.5) --(\x-0.9,0.5);
\draw [thick] (\x-2,-0.5)--(\x-0.9,-0.5);
\draw [thick] (\x+0.9,0.5) --(\x+2.1,0.5);
\draw [thick] (\x+0.9,-0.5)--(\x+2.1,-0.5);
\draw [thick] (\x+1,0)--(\x+2,0);
\draw [thick] (\x-0.4,0) --(\x+0.4,0);
\draw [thick] (\x,-0.4) --(\x,0.4);
\def \x {25}
\draw [thick] (\x,0) circle (1);
\draw [thick] (\x+0.9,0.5) --(\x+2,0.5);
\draw [thick] (\x+0.9,-0.5)--(\x+2,-0.5);
\draw [thick] (\x-0.4,0) --(\x+0.4,0);
\end{tikzpicture}
\end{center}
\vskip-43pt\be\label{s4}\phantom{x}\ee\vskip6pt
in `bubble' notation. The threshold at which this process becomes possible corresponds to another branch point at $s = (m_a+m_b+m_c)^2$, and [\ref{s4}] gives the total discontinuity across the two cuts along the real axis. And so, as and when new processes become possible at higher energies, the appearance of corresponding additional terms in the unitarity equations implies a sequence of branch point singularities, termed normal thresholds. 

The unitarity equations can also imply the presence of poles, as well as cuts, in the physical region, {\it i.e.} for real, physical values of the momenta. At suitable values of momenta, the unitarity equation for three-to-three scattering, $a+b+c\rightarrow a+b+c$, the unitarity equation contains a term
\begin{center}
\begin{tikzpicture}[scale=0.5]
\def \x {0}
\def \y {0}
\node [left] at (-2,0.5) {$b$};
\node [left] at (-2,-0.5) {$c$};
\node [right] at (4,-0.5) {$c$};
\draw [thick] (\x,\y) circle (1);
\draw [thick] (\x-2, \y+0.5) --(\x-0.9,\y+0.5);
\draw [thick] (\x-2,\y-0.5)--(\x-0.9,\y-0.5);
\draw [thick] (\x+0.9, \y-0.5)--(\x+4,\y-0.5);
\draw [thick] (\x-0.4,\y) --(\x+0.4,\y);
\draw [thick] (\x,\y-0.4) --(\x,\y+0.4);
\draw [thick] (\x+0.71,\y+0.71) --(\x+2-0.71,\y+2-0.71);
\def \x {2}
\def \y {2}
\node [left] at (-2,2.5) {$a$};
\node [right] at (4,2.5) {$a$};
\node [right] at (4,1.5) {$b$};
\draw [thick] (\x,\y) circle (1);
\draw [thick] (\x-4, \y+0.5) --(\x-0.9,\y+0.5);
\draw [thick] (\x+0.9, \y+0.5) --(\x+2, \y+0.5);
\draw [thick] (\x+0.9, \y-0.5)--(\x+2,\y-0.5);
\draw [thick] (\x+0.4,\y) --(\x-0.4,\y);
\end{tikzpicture}
\end{center}
\vskip-60pt\be\label{s5}\phantom{x}\ee\vskip26pt
which corresponds to a $\delta(q^2-m_b^2)$ discontinuity in the amplitude, where $q=p_a+p_b-p_a'$, where $p_a,p_b$ denote incoming momenta, and $p_a'$ an outgoing momentum, for the appropriate particles. David argued that this implies that the amplitude, viewed as an analytic function, has a pole at $q^2=m_b^2 $. 
\vspace{-5truemm}
\begin{center}
\begin{tikzpicture}[scale=0.5]
\def \x {0}
\node [left] at (\x-2.7,0.9) {$a$};
\node [left] at (\x-2.7,0) {$b$};
\node [left] at (\x-2.7,-0.9) {$c$};
\node [right] at (\x+2.7,0.9) {$a$};
\node [right] at (\x+2.7,0) {$b$};
\node [right] at (\x+2.7,-0.9) {$c$};
\draw [thick] (\x,0) circle (1.5);
\draw [thick] (\x-2.7,0.9) --(\x-1.2,0.9);
\draw [thick] (\x-2.7,0) --(\x-1.5,0);
\draw [thick] (\x-2.7,-0.9)--(\x-1.2,-0.9);
\draw [thick] (\x+1.2,0.9) --(\x+2.7,0.9);
\draw [thick] (\x+1.5,0) --(\x+2.7,0);
\draw [thick] (\x+1.2,-0.9)--(\x+2.7,-0.9);
\draw [thick] (\x-0.4,0) --(\x+0.4,0);
\draw [thick] (\x,-0.4) --(\x,0.4);
\node at (5.5,0) {\Large${\sim }$};
\def \x {10}
\def \y {-1}
\node [left] at (\x-2,\y+0.45) {$b$};
\node [left] at (\x-2,\y-0.45) {$c$};
\node [right] at (\x+4,\y-0.45) {$c$};
\draw [thick] (\x,\y) circle (1);
\draw [thick] (\x-2, \y+0.45) --(\x-0.9,\y+0.45);
\draw [thick] (\x-2,\y-0.45)--(\x-0.9,\y-0.45);
\draw [thick] (\x+0.9, \y-0.45)--(\x+4,\y-0.45);
\draw [thick] (\x-0.4,\y) --(\x+0.4,\y);
\draw [thick] (\x,\y-0.4) --(\x,\y+0.4);
\draw [thick] (\x+0.71,\y+0.71) --(\x+2-0.71,\y+2-0.71);
 \draw [thick] (\x+1,\y+1) circle (0.15);
\def \x {12}
\def \y {1}
\node [left] at (\x-4,\y+0.45) {$a$};
\node [right] at (\x+2, \y+0.45) {$a$};
\node [right] at (\x+2,\y-0.45){$b$};
\draw [thick] (\x,\y) circle (1);
\draw [thick] (\x-4, \y+0.45) --(\x-0.9,\y+0.45);
\draw [thick] (\x+0.9, \y+0.45) --(\x+2, \y+0.45);
\draw [thick] (\x+0.9, \y-0.45)--(\x+2,\y-0.45);
\draw [thick] (\x+0.4,\y) --(\x-0.4,\y);
\draw [thick] (\x,\y-0.4) --(\x,\y+0.4);
\node at (20,0) {at $q^2=m_b^2$\,,};
\end{tikzpicture}
\end{center}
\vskip-56pt\be\label{s6}\phantom{x}\ee\vskip16pt
where the middle line on the right hand side denotes a factor of  $(q^2-m_b^2)^{-1}$, rather than a $\delta$-function. 

The unitarity equation for the  $a+b+c\rightarrow a+b+c$
scattering amplitude also contains the term (a) shown in [\ref{s7}]. If the pole in [\ref{s6}] is inserted in this term, the structure corresponding to the term (b) in [\ref{s7}] is obtained [(11);(12), pp. 206, 266--278].
\begin{center} 
\begin{tikzpicture}[scale=0.5]
\node at (-4.5,0) {(a)};
\def \x {0}
\draw [thick] (\x,0) circle (1.5);
\draw [thick] (\x-2.7,0.9) --(\x-1.2,0.9);
\draw [thick] (\x-2.7,0) --(\x-1.5,0);
\draw [thick] (\x-2.7,-0.9)--(\x-1.2,-0.9);
\draw [thick] (\x+1.2,0.9) --(\x+5.6,0.9);
\draw [thick] (\x+1.5,0) --(\x+2.6,0);
\draw [thick] (\x+1.2,-0.9)--(\x+2.6,-0.9);
\draw [thick] (\x-0.4,0) --(\x+0.4,0);
\draw [thick] (\x,-0.4) --(\x,0.4);
\draw [thick] (\x+3.5,-0.5) circle (1);
\draw [thick] (\x+4.4,0) --(\x+5.6,0);
\draw [thick] (\x+4.4,-0.9)--(\x+5.6,-0.9);
\draw [thick] (\x+3.1,-0.5) --(\x+3.9,-0.5);
\node at (9.2,0) {(b)};
\def \x {13}
\def \y {-1}
\draw [thick] (\x,\y) circle (1);
\draw [thick] (\x-2, \y+0.45) --(\x-0.9,\y+0.45);
\draw [thick] (\x-2,\y-0.45)--(\x-0.9,\y-0.45);
\draw [thick] (\x+0.9, \y-0.45)--(\x+3,\y-0.45);
\draw [thick] (\x-0.4,\y) --(\x+0.4,\y);
\draw [thick] (\x,\y-0.4) --(\x,\y+0.4);
\draw [thick] (\x+0.71,\y+0.71) --(\x+2-0.71,\y+2-0.71);
 \draw [thick] (\x+1,\y+1) circle (0.15);
\def \x {15}
\def \y {1}
\draw [thick] (\x,\y) circle (1);
\draw [thick] (\x-4, \y+0.45) --(\x-0.9,\y+0.45);
\draw [thick] (\x+0.9, \y+0.45) --(\x+4, \y+0.45);
\draw [thick] (\x+0.4,\y) --(\x-0.4,\y);
\draw [thick] (\x,\y-0.4) --(\x,\y+0.4);
\def \x {17}
\def \y {-1}
\draw [thick] (\x-0.71,\y+0.71) --(\x-2+0.71,\y+2-0.71);
\draw [thick] (\x,\y) circle (1);
\draw [thick] (\x-2,\y-0.45)--(\x-0.9,\y-0.45);
\draw [thick] (\x+0.9, \y-0.45)--(\x+2,\y-0.45);
\draw [thick] (\x+0.9, \y+0.45)--(\x+2,\y+0.45);
\draw [thick] (\x-0.4,\y) --(\x+0.4,\y);
\end{tikzpicture}
\end{center}
\vskip-56pt\be\label{s7}\phantom{x}\ee\vskip26pt
Related terms are generated by other terms in the unitarity equation and together they imply a cut in $(s,t)$-plane with discontinuity given by
\begin{center}
\begin{tikzpicture}[scale=0.5]
\def \x {13}
\def \y {-1}
\draw [thick] (\x,\y) circle (1);
\draw [thick] (\x-2, \y+0.45) --(\x-0.9,\y+0.45);
\draw [thick] (\x-2,\y-0.45)--(\x-0.9,\y-0.45);
\draw [thick] (\x+0.9, \y-0.45)--(\x+3,\y-0.45);
\draw [thick] (\x-0.4,\y) --(\x+0.4,\y);
\draw [thick] (\x,\y-0.4) --(\x,\y+0.4);
\draw [thick] (\x+0.71,\y+0.71) --(\x+2-0.71,\y+2-0.71);
\def \x {15}
\def \y {1}
\draw [thick] (\x,\y) circle (1);
\draw [thick] (\x-4, \y+0.45) --(\x-0.9,\y+0.45);
\draw [thick] (\x+0.9, \y+0.45) --(\x+4, \y+0.45);
\draw [thick] (\x+0.4,\y) --(\x-0.4,\y);
\draw [thick] (\x,\y-0.4) --(\x,\y+0.4);
\def \x {17}
\def \y {-1}
\draw [thick] (\x-0.71,\y+0.71) --(\x-2+0.71,\y+2-0.71);
\draw [thick] (\x,\y) circle (1);
\draw [thick] (\x-2,\y-0.45)--(\x-0.9,\y-0.45);
\draw [thick] (\x+0.9, \y-0.45)--(\x+2,\y-0.45);
\draw [thick] (\x+0.9, \y+0.45)--(\x+2,\y+0.45);
\draw [thick] (\x-0.4,\y) --(\x+0.4,\y);
\draw [thick] (\x,\y-0.4) --(\x,\y+0.4);
\end{tikzpicture}
\end{center}
\vskip-56pt\be\label{s8}\phantom{x}\ee\vskip26pt
just as in perturbative quantum field theory (QFT). In a series of three papers (15--17) with his student, Michael Bloxham, and John Polkinghorne, David{} was able to extend this approach to show that for physical values of the momenta, unitarity requires scattering amplitudes to be singular on the arcs of curves where Landau (1959) had determined that singularities occur for Feynman diagrams in QFT, and that the discontinuities associated with these singularities are given by the rules obtained for perturbative QFT by Cutkosky (1960).

Building on ideas of Gunson (later published in Gunson (1965); see p. 847), David also showed how similar arguments based on analyticity and unitarity imply the existence of antiparticles, and may allow the proof of crossing symmetry and the TCP theorem (5,12), as well as implying the standard connection between spin and statistics (8) [{\it i.e.}~that integral spin particles are bosons and half-odd-integral ones are fermions]. However, these results depend on the existence of suitable paths for analytic continuation and a knowledge of the singularity structure of the $S$-matrix outside the physical region and for complex values of the momenta.  

Reviewing the state of $S$-matrix theory nearly a decade after he ceased working on it, David reiterated his belief that ``Properties involving analyticity well away from [the physical region] should surely not be postulated but rather deduced from more
fundamental principles involving the physical region.'' The developments of the early 1960s, in which David had played a leading role, had shown how one might hope to re-establish the main achievements of Axiomatic QFT (see Streater and Wightman 1964) within the (arguably) more general framework of $S$-matrix theory, but, as David put it, 
``Although the general
features of the physical region were appreciated, a detailed, precise and
comprehensive mathematical treatment was lacking.''(48)
Later approaches to providing such a mathematical treatment for discussing the $S$-matrix near the physical region were developed by Sato (1975) and by Iagolnitzer (1981), but no progress on establishing a  rigorous analysis of non-physical region singularity structure  has been made.

\bigskip\bigskip
\bigskip
\centerline{\sc Fermions and the GSO Projection}
\bigskip
\nobreak
At early in 1971, Pierre Ramond proposed a description of  free fermions in dual models (Ramond 1971) in a sort of generalization of the Dirac equation. This was quickly extended by Andr\'e Neveu and John Schwarz who gave expressions for dual model amplitudes involving a single Ramond fermion interacting with a sequence of mesons (Neveu \& Schwarz 1971). When he arrived at CERN some months later to take up his staff appointment, David began working with Edward Corrigan on a programme to construct a complete consistent theory of dual fermions and mesons working within the operator formalism.  

The operator formalism of the original bosonic dual model of Veneziano, the space of states is created by bosonic oscillators, $a_n^\mu$, where
$\mu$ runs over the dimensions of space-time and $n$ runs over the  integers. Neveu and Schwarz enlarged this using anti-commuting
 fermionic oscillators, $b_r^\mu$,  where $r$ 
runs over half odd integers, in addition to the $a_n^\mu$, to create a space of space-time boson states, while the Ramond states, which are  space-time fermions, are created by fermionic oscillators, $d^\mu_n$, where $n$ runs over the  integers, in addition to the $a_n^\mu$. These oscillators satisfy the commutation and anti-commutation relations,
 \be\label{f1}
[a_m^\mu, a_n^\nu ] = m\delta_{m,-n}\eta^{\mu\nu},\qquad
\{b_r^\mu, b_s^\nu \} = \delta_{r,-s}\eta^{\mu\nu},\qquad \{d_m^\mu, d_n^\nu \} = \delta_{m,-n},\eta^{\mu\nu},
 \ee
where $\eta^{\mu\nu}$ is the space-time metric, 
and
$ a_m^{\mu\dagger} = a_{-m}^\mu, \,b_r^{\mu\dagger}=b_{-r}^\mu,
\,d_n^{\mu\dagger}=d_{-n}^\mu$.

The Neveu-Schwarz states are created by the action of the oscillators  $a_m^\mu, b_r^\mu, $ on a vacuum state 
 $|0\rangle$ that is annihilated by the  operators $a_m^\mu, m>0,$
 and $b_r^\mu, r>0.$ The Fock space created in this way has a natural scalar product
which is not positive because the space-time metric $\eta^{\mu\nu}$ is not. The consistency of a dual model, such as the original Veneziano model or the Ramond-Neveu-Schwarz model, requires that the physical states that couple in amplitudes belong to a positive definite subspace defined by an infinite set of gauge conditions, {\it i.e.} that there are no `ghost' physical states. 

The first step in constructing a complete dual theory of fermions and bosons was to rewrite the amplitudes of Neveu and Schwarz in a dual form corresponding to a boson annihilating into a fermion pair. In essence, this required the construction of a `fermion emission vertex', an operator describing the process by which a fermion changes into a boson by the emission of a fermion. Such a construction had been found by Thorn (1971) and by Schwarz (1971). Corrigan and Olive (28) gave this a more elegant and manageable formulation, making the action of the vertex on the gauge conditions more transparent. A prime objective was the calculation of the amplitude for fermion-fermion scattering by combining a fermion emission vertex and its conjugate.

The defining properties of  this vertex operator,  $W_\chi(z)$, as formulated by Corrigan and Olive, are that it maps boson (or meson) states into fermion states in such a way that it intertwines between a Neveu-Schwarz field, $H^\mu(y) =\sum b_r^\mu \,y^{-r}$ and a Ramond field, $\Gamma^\mu(y) = \sum d^\mu_n\,y^{-n}$, according to an equation of the form,
\be \label{f2}
W_\chi(z){H^\mu(y)\over\sqrt{y}}=\lambda
{\Gamma^\mu(y-z)\over\sqrt{y-z}} W_\chi (z), 
\ee
where $\lambda$ is a suitable constant,
and that it creates the fermion state, $|\chi\rangle$, from the boson vacuum,  $\vac$: $W_\chi(z)\vac=e^{zL_{-1}}|\chi\rangle$. 

Whereas the vertices describing the emission of bosons essentially commute with the gauge conditions that eliminate ghost states, the behaviour of the fermion emission vertex is more complicated. It converts an infinite linear combination of gauge conditions in the boson sector into an infinite combination in the fermion sector. It did not immediately follow that ghost states would not couple in the four-point fermion amplitude.  

Working with Lars Brink, David saw that it was necessary to introduce a projection operator onto the space of physical states to secure this (30). As a preparation, Brink and Olive looked at the algebraically related problem of calculating the one loop contribution in the original bosonic theory, giving a precise derivation (31) of the results earlier proposed heuristically by Lovelace (1971), that had first suggested that the space-time in bosonic string theory needed to be 26 for consistency. Next, with Claudio Rebbi and Jo\"el Scherk, they worked  out exactly how the
gauge conditions related to the fermion emission vertex,  
under
the requirement that the lowest fermion mass $m=0$ and the dimension of space-time $D=10$,
and found that the calculation of the four-point fermion amplitude was surprising similar to the one loop calculation. 

Using these results,  Olive and Scherk (35) were able to calculate that the effect of projecting onto the physical states satisfying the gauge conditions was to introduce a factor $1/\Delta(x)$
into the integral expression for the four-point fermion amplitude. The function $\Delta(x)$ was defined as an infinite determinant which was not readily evaluated. After John Schwarz and Cheng-Chin Wu (1974) using a computer calculation to propose $\Delta(x)=(1-x)^{1\over 4}$, a result that was proved analytically soon after by Corrigan, Goddard, Olive and Russell Smith (36). Meanwhile, Stanley Mandelstam (1973) had obtained the same result very efficiently using his mastery of the light-cone formulation of string theory.

The conclusion of the calculation of the four fermion amplitude not only in the $s$ channel, where, by construction, the poles correspond to the propagation of physical states, but also in the $t$ channel, where the poles correspond to the same spectrum. Before the calculation, it was not certain not that the amplitude would be meromorphic: conceivably it could have had cuts in the $t$ channel. There was however a subtle discrepancy, which proved to be an unappreciated clue to a deeper structure: the parity of the boson states was changed between the $s$ and $t$ channels, so that if the lowest mass state in the $s$ channel, a tachyon, is a pseudo-scalar, the lowest mass state in the $t$ channel is a scalar tachyon. 

The structure of the four fermion amplitude gave much encouragement that a consistent Ramond-Neveu-Schwarz model could be constructed, but the parity doubling was puzzling. By 1974, it had been realized by various people that it was possible to make a projection, consistent with interactions, that would remove something like half the states including the tachyon in the bosonic sector, and leave a massless chiral fermion as the lowest state in the fermionic sector.   
In 1974, the focus of David's interest shifted towards the study of monopoles in gauge theories, but in 1976 he was drawn back to dual fermion theories for what proved to be a very seminal interlude. 

In the summer of 1975, David began thinking about the behaviour of spinors in different dimensions. A spinor in even dimension $D$ of space-time in general has $2^{D/2}$ components. This can be halved by imposing either of two conditions; that the spinor be Weyl (chiral) or that it be Majorana (real). These conditions are incompatible unless  $D=2$ (mod 8). David was struck by the fact that $D=10$, the dimension appropriate for the Ramond-Neveu-Schwarz model, is the smallest non-trivial dimension in which there are Majorana-Weyl spinors. Such spinors have  $2^{10/2-2}=8$
components.

In 1976, Jo\"el Scherk, who had returned to the \'Ecole  Normale Sup\'erieure in Paris, and Fernando Gliozzi, who was visiting, had begun studying the multiplicity of the states at various mass levels in the Ramond-Neveu-Schwarz model. They noticed, first by manual calculation, there was a remarkable coincidence between the number of states at a given mass level in the bosonic sector and the same level in the fermionic sector, provided that they could impose both Weyl and Majorana conditions. Such a projection would also eliminate the problem with parity doubling that had surfaced in the four fermion amplitude. 

Soon they discovered that there was an identity in the literature, proved by Jacobi in 1829,
\be \label{f3}
{1\over 2w^\half}\left[\prod_{m=1}^\infty\left({1+w^{m-\half}\over 1-w^m}\right)^8
-\prod_{m=1}^\infty\left({1-w^{m-\half}\over 1-w^m}\right)^8
\right]
=8\prod_{m=1}^\infty\left({1+w^{m}\over 1-w^m}\right)^8\,,
\ee
that guarantees equal numbers of bosons and fermions at each mass level.
The coefficient of $w^n$ on the left hand side of this relation gives the number of bosonic states at the $n$-th mass level (above $m=0$) and the right hand side does the same for the fermionic sector, provided that the spinors are subject to two conditions. 
From the beginning, two-dimensional supersymmetry, on the world sheet of the string, had been at the heart of the structure of the Ramond-Neveu-Schwarz model, but the equality of the numbers of bosons at each mass level strongly suggested that the theory was also supersymmetric in space-time, and this was subsequently shown to be the case.

When Jo\"el Scherk visited CERN, to give a talk on developments in supergravity, he told David of these results and David explained to Jo\"el that the space-time dimension being 10 plays an essential role in being able to impose both Majorana and Weyl conditions, necessary for the factor 8 in [\ref{f3}]. Because of the importance of this insight, Jo\"el invited David to join the collaboration, and the projection has become known as the Gliozzi-Scherk-Olive, or GSO, projection (44,45), and the theory it defined is what has become known as superstring theory.  For the first time, there was a perturbatively unitary, tachyon free, space-time supersymmetric dual theory.

\bigskip
\bigskip
\bigskip
\centerline{\sc Monopoles and Duality}
\bigskip
\nobreak
In one of the most breathtakingly original contributions he made in the early years of Quantum Mechanics, Paul Dirac (1931) showed that the consistency of quantum 
mechanics for a system involving both electric charges, $q$, and magnetic charges, $g$, requires that each pair of such charges satisfy
\be
{qg\over 4\pi\hbar}={n\over 2},\qquad n\hbox{ an integer}.\label{m1}
\ee
This relation implies that the striking conclusion that the existence of a single magnetic charge requires the existence of a smallest electric charge, $q_0$, with any other electric charge being an integral multiple of $q_0$, and, similarly, that there is a smallest magnetic charge,  $g_0$, of which any other magnetic charge must be an integral multiple.

Dirac was discussing point charges, singularities in the electromagnetic field, but, the mid 1970s, in nearly simultaneous papers, 't Hooft (1974) and Polyakov (1974) showed that extended objects with magnetic charge can occur as smooth classical solutions in certain familiar gauge field theories. They considered an SO(3) gauge theory with an adjoint representation ({\it i.e.}~triplet real) scalar field $\bphi$, whose self-interactions are described by the Higgs potential,
\be
V(\bphi)= {1\over 4}  \lambda ( \bphi^2-a^2)^2 .\label{m2}
\ee
 In the lowest energy state $\bphi =\bphi_0,$ constant, a value on the surface $\M$ of minima of $V$, which is the sphere $\bphi^2=1$. All such choices of $\bphi_0$ are gauge-equivalent and each choice corresponds to a spontaneous breaking of the gauge group down to U(1), which can be identified with the gauge group of electromagnetism.

't Hooft and Polyakov showed that the classical field equations of this SU(2) Higgs model have a smooth spherically symmetric solution under simultaneous rotations in ordinary space and the three-dimensional space in which the Higgs field $\bphi$ lives. This implies making a particular identification between directions in these spaces; with this,  $\bphi$ is everywhere radial, vanishing at the origin. Asymptotically, $\bphi(\br)\sim a\hat\br,$ and its U(1) little group is identified with the electromagnetic gauge group. The corresponding component of the gauge field strength gives the electromagnetic field, which, for the 'tHooft-Polyakov solution is a purely magnetic field, $\bB\sim (1/er^2)\hat\br$ asymptotically, where $e$ is the gauge coupling constant; this is the field of a magnetic monopole of strength,
\be
g={4\pi\over e}.
\label{m3}
\ee
In the quantized SU(2) gauge theory, the electric charges $q$ are multiples of $\half \hbar e$ and so such charges satisfy the Dirac quantization condition [\ref{m1}] with the magnetic charge $g$ given by [\ref{m3}].

To understand the 't Hooft-Polyakov monopole more deeply, David began to investigate whether similar solutions in gauge theories might provide theories of particles, such as hadrons, working with Edward Corrigan, David Fairlie and Jean Nuyts (40). They sought generalizations of it by looking for solutions to an SU(3) Higgs model that were spherically symmetric in an appropriate sense.  In the case of SU(3), for a generic quartic (and so renormalizable) Higgs potential, $V$, the gauge group acts transitively on the vacuum manifold, $\M$, of minima of $V$, and the unbroken symmetry group, the little group of the Higgs field, $\bphi$, is $\SU(2)\times\U(1)/\Zop_2$, and we can again identify the U(1) factor with electromagnetism. 

The mapping between ordinary space and the space of the Higgs field, in the case of the 't Hooft-Polyakov monopole, determines an isomorphism between SU(2), the covering group of the group of spatial rotations SO(3), and the SU(2) gauge group. For other gauge groups, such as SU(3), given a homomorphism of SU(2) into the gauge group, spherical symmetry can be defined as invariance under the simultaneous applications of rotations and the corresponding gauge transformations under the homomorphism. If $t^i, i=1,2,3,$ are the images of a standard basis for the generators of SU(2) in the Lie algebra of the gauge group, spherical symmetry corresponds to invariance under transformations generated by $-i\br\wedge\nabla+\bt$.

David and his collaborators noted that there are two essentially distinct homomorphisms of SU(2) into SU(3), that is ones not related by SU(3) gauge transformation: (i) one in which the image is SU(2); and (ii) another in which the image is SO(3). They found spherically symmetric solutions, possessing U(1) magnetic charge, for each of these cases, but, while in case (ii) the Dirac condition [\ref{m1}] was satisfied, for case (i) there were solutions for which the condition had to be relaxed to allow $n$ to be half-integral. These solutions evaded the original  Dirac condition by having a long range magnetic type SU(2) gauge field, not part of the context of Dirac's original argument. 

In his next contribution, with Edward Corrigan (42), David derived a generalization of the Dirac condition which encompassed such cases. They considered a theory with gauge group $G$ spontaneously broken, {\it e.g.} by an adjoint representation Higgs field, to a subgroup $H=K\times\U(1)/Z$, where $Z=K\cap\U(1)$ is discrete; $K$ can be identified with a `colour' gauge group and U(1) with electromagnetism. They showed that, for any smooth solution, the magnetic charge $g$ associated with the U(1) factor satisfies the generalized Dirac condition,
\be
\exp(igQ)\in K\,,
\label{m4}
\ee
where $Q$ is the electric charge operator, which generates the U(1). If $k$, the element of $K$ defined by [\ref{m4}], equals $1$, then this condition is equivalent to the original Dirac condition  [\ref{m1}] and this is the case for $G = \SU(2)$, where $K$ is trivial. For $G=\SU(3)$, in case (ii) $k=1$, but in case (i) $k\ne 1$, which is why the condition  [\ref{m1}] can be violated in this case.  Corrigan and Olive noted that, if $k\ne 1$, the generalized condition [\ref{m4}] implies that [\ref{m1}]  holds for the electric charges of colour singlet particles, establishing a link between colour and fractional electric charges. 

For spherically symmetric monopole solutions, David showed (43) that the image, in the Lie algebra of $G$, of the generator of rotations about the radial direction
$\hat\br\cdot\bt=-gQ+\kappa$, where $\kappa$ is a generator of $K$. The condition [\ref{m3}] follows again from this relation because the eigenvalues of $\hat\br\cdot\bt$ are half integral. Spherical symmetry implies that $\hat\br\cdot\bt$ is a generator of the little group, $H$, of $\bphi(\br)$ and so $\, h(s)=\exp(4\pi is\hat\br\cdot\bt ), \, 0\leq s\leq 1,$ defines a closed loop in $H$ and an element of the homotopy group, $\Pi_1(H)$, which David show equalled the topological invariant or `topological quantum number' (although it is a classical concept), which had earlier been associated with the solution (Tyupkin {\it et al.}~1975, Monastyrski\u i \& Perelomov 1975). The asymptotic limit of the Higgs field, $\bphi_\infty(\hat\br)
=\lim_{r\rightarrow\infty}\bphi(r\hat\br)$, defines a map $S^2\rightarrow\M\cong G/H$, and so an element of the homotopy group $\Pi_2(G/H)$, which is isomorphic to 
$\Pi_1(H)$ (assuming $G$ to be simply connected). David showed that the element of $\Pi_1(H)$ associated with the solution in this way corresponds to the loop $h$.

Seeking to deepen our understanding of magnetic monopoles in gauge theories, working with Peter Goddard and Jean Nuyts  (GNO), David next investigated (46) the general case of a theory with a compact gauge group $G$, spontaneously broken to an exact gauge group $H\subset G$ by a Higgs field, $\bphi$, in an arbitrary representation. They considered monopole solutions, that is solutions for which the spatial components of the gauge field strength $\G_{\alpha\beta}$, expressed as a generator of $H$, the little group of $\bphi_\infty(\hat\br)$, the asymptotic value of $\bphi(\br)$ in the radial direction, 
\be
\G_{ij}(\br)\sim {\epsilon_{ijk}r_k\over r^3}\;\G(\hat\br)\,,\qquad\qquad\hbox{as } r\rightarrow\infty\,,\quad 1\leq i,j,k\leq 3\,.
\label{m5}
\ee
Generalizing [\ref{m1}] and [\ref{m4}], they established that 
\be
\exp(ie\,\G)=1\,, 
\label{m6}
\ee
where $\G=\G(\hat\br)$, and that the loop 
$\, h(s)=\exp(ise\G), \, 0\leq s\leq 1,$ determines the topological invariant associated with the solution as an element of $\Pi_1(H)$. Taking a maximal set of commuting generators of $H$, $T^a, 1\leq a\leq m,$ where $m$ is the rank of $H$, {\it i.e.} a basis for a Cartan subalgebra of $H$, $\G$ will be equivalent, under an $H$ gauge transformation, to a linear combination, $\G=g_aT^a$, of the $\{T^a\}$. $\G$ is the generalized magnetic charge of the solution, and GNO called  $\bg=(g_a)$ its magnetic weights. 

The simultaneous eigenvalues $\bq=(q^a)$  of  $\hbar eT^a, 1\leq a\leq m,$ in representations of $H$, are the generalized electric charges of quanta fields, transforming under those representations. The quantization condition [\ref{m6}] can be re-expressed in the form
\be
{\bq\cdot\bg\over 4\pi\hbar}={n\over 2}\,,\qquad n\hbox{ an integer},\label{m7}
\ee
more closely paralleling [\ref{m1}]. The possible values of $\bq$ are points of the lattice $\hbar e\Lambda_H$, where $\Lambda_H$ is the weight lattice of the group $H$, implying that $\bg\in(2\pi/e){\Lambda_H}^\ast$, where ${\Lambda_H}^\ast$ is the lattice dual to $\Lambda_H $. The weight lattice, $\Lambda_{\widetilde H} $, of weights of the simply connected group, 
$\widetilde H$, with the same Lie algebra as $H$, consists of those $\blambda$ for which $2\balpha\cdot \blambda/\balpha^2$ is an integer for every root $\balpha$ of $H$. This condition is satisfied by the roots themselves, and thus $\Lambda_R\subset \Lambda_H \subset \Lambda_{\widetilde H}$, where $\Lambda_R$ is the root lattice of $H$, the lattice generated by its roots, $\balpha$. If $H$ is a simple group all of whose roots have the same length ({\it i.e.} $H$ is simply-laced), we can normalize so that  $\balpha^2=2$ for each root, and then $\Lambda_R\subset {\Lambda_H}^\ast \subset \Lambda_{\widetilde H}$. Thus ${\Lambda_H}^\ast $ is the weight lattice of some group, $H^\vee$ say, called the dual of $H$, with the same Lie algebra as $H$ but a different global structure in general, {\it e.g.} $\SU(2)^\vee=\SO(3)$ and conversely. The relationship between a group and its dual is reflexive:  $(H^\vee)^\vee=H$.

If $H$ is not simply-laced, ${\Lambda_H}^\ast $ is still the weight lattice of some group $H^\vee$, the dual of $H$, but now it is a group with roots $\balpha^\vee=2\balpha/\balpha^2$, where $\balpha$ is a root of $H$, and its algebra may not be isomorphic to that of $H$. The simple Lie algebras that are not simply-laced are the series $\hbox{B}_n$ and $\hbox{C}_n$, corresponding to, {\it e.g.},~the groups SO($2n+1$) and Sp($n$), respectively, and the exceptional algebras $G_2$ and $F_4$. For the groups with Lie algebras $G_2$ and $F_4$, the dual groups have the same Lie algebra, but different global structure. However, the dual of a group with Lie algebra $\hbox{B}_n$ is one with Lie algebra $\hbox{C}_n$, and {\it vice versa}.

Thus, GNO showed that, while the generalized electric charges are, up to a factor, the weights of the exact symmetry group, $H$, the generalized magnetic charges of the extended monopole solutions are, up to a factor,  weights of the dual group $H^\vee$. They suggested that these solutions should form multiplets of $H^\vee$ and that this group should be an exact symmetry group of the theory as well as $H$. Further, they speculated that the 
the dual relationship between the generalized magnetic charges, $\bg$, associated with the topological characteristics of the monopole solution and the group $H^\vee$, on the one hand, and the generalized electric charges, $\bq$, associated with the quanta of the basic fields of the theory and the group $H$, on the other, was analogous to that between the solitons and the quanta of the basic boson field in the two-dimensional sine-Gordon theory.  

GNO noted that the sine-Gordon theory had been shown to be equivalent to the massive Thirring model at the quantum level, with the solitons, which are extended solutions in the former boson theory, corresponding to the quanta of the basic fermion field in the Thirring model (Coleman 1975, Mandelstam 1975), ideas that go back to the work of T. H. R. Skyrme (Skyrme 1961, and references therein). In this correspondence, the topologically conserved soliton number in the sine-Gordon theory corresponds to a conserved Noether charge associated with a U(1) symmetry of the Thirring model. But this exact equivalence of different theories is intrinsically quantum mechanical and so any such equivalence between dual theories, interchanging electric and magnetic charges, is beyond the basically classical analysis of GNO (46). However, they speculated that the monopoles of the original theory might correspond to particles in a fundamental representation of $H^\vee$ in a dual formulation; in his next investigations, David was led instead to a more elegant and profound conjecture.

David was determined to explore further what he acknowledged were highly speculative propositions, that $H$ monopoles behave as irreducible representations of $H^\vee$, and that the theory has a $H^\vee$ gauge symmetry. Working with Claus Montonen (47), he looked for evidence for the conjectures in the simplest monopole theory, that studied by 't Hooft and Polyakov, in which the gauge group $G=\hbox{SO}(3)$ is spontaneously broken to $H=\hbox{SO}(2)\cong\hbox{U}(1) $ by the choice of a particular vacuum value for $\bphi$ minimizing the Higgs potential [\ref{m2}]. The mass, $M_M$, of the monopole solutions in this theory is of the form,
\be
M_M = {4\pi a\over e}f(\lambda/e^2), \label{m8}
\ee
where $f(\lambda/e^2)\rightarrow 1$ as 
$\lambda\rightarrow 0$ and following Prasad and Sommerfield (1975), Montonen and Olive considered the limiting value, $\lambda=0$, in which the potential vanishes but the asymptotic condition $\bphi^2\rightarrow a^2$ is maintained at spatial infinity. Then, from [\ref{m3}], $M_M=ag$, which is of exactly the same form as the formula for the massive of the massive spin 1 particles, $W^\pm$, generated by the Higgs mechanism, namely $M_W= ae\hbar =aq$. Bogomol'nyi (1976) showed that, for general $\lambda$, $M_M\geq ag $, so that the monopole mass reaches its lower bound at  $\lambda=0$, which is now known as the Bogomol'nyi-Prasad-Sommerfield (BPS) limit.  

Montonen and Olive conjectured that there should be dual equivalent versions of this theory, described by Lagrangians of the same form. Under this duality correspondence, electric and magnetic properties interchange, the massive `gauge' particles,  $W^\pm$, exchange r\^oles with monopole solutions, and   
 topological and Noether quantum numbers are interchanged. In addition to the symmetry between the mass formulae  $M_W =aq$ and $ M_M=ag $, they noted that further circumstantial evidence for the duality conjecture was provided by Manton's calculation of the classical magnetic force between monopoles as $g^2/4\pi r^2$, symmetric with the electric case (Manton 1977).
 
Because [\ref{m3}] implies that as $e$ becomes small, $g$ becomes large, this Montonen-Olive duality provides a (conjectured) equivalence between theories at strong and weak coupling. It was the first example of what became known as S-duality, which make a central concept in quantum field theory and string theory from the mid 1990s onwards. 

Notwithstanding the suggestive evidence in favour of the extremely bold  Montonen-Olive conjecture, there were significant potential obstacles to its validity: Why should the symmetry between the formulae for the masses of monopoles and  massive gauge particles survive renormalization? Why should the monopole have spin 1 quantum mechanically?  Answers to these two questions came quickly. With Edward Witten, David showed  that in suitable $N=2$ supersymmetric theories, the presence of monopole solutions results in the existence of central terms that modify  the supersymmetry algebra, with the mass formulae following as a consequence of the algebra (50). Then Hugh Osborn (1979) observed that in $N=4$ supersymmetric gauge theory, with spontaneous symmetry  breaking, the monopole states would have spin 1 like the massive gauge particles, making it the most suitable theory for realizing the Montonen-Olive duality conjecture. Fifteen years later, Nathan Seiberg and Edward Witten (1994a,b) found a form of S-duality that worked in $N=2$ supersymmetric theories.

David continued to study the characteristics of monopole solutions to gauge theories, motivated in large part by his duality conjecture. With Peter Goddard, he studied theories in a gauge group $G$ is broken by an adjoint representation Higgs field to an exact symmetry gauge group, $H$, whose structure is locally of the form $\hbox{U}(1)\times K$ (60). They found that a necessary and sufficient condition for $K$ to be semi-simple is that the vacuum value of the Higgs field be on the same orbit of $G$ as a fundamental weight, and they gave prescriptions for determining $K$, and for calculating the basic units of electric and magnetic charge, from the Dynkin diagram of $G$. They also showed that the global structure of $H = (\hbox{U}(1)\times  K)/Z $, where $Z$ is a cyclic group whose order can also be calculated from the Dynkin diagram.

Next (61), they considered such theories in the BPS limit, analysing the charges of potentially stable magnetic monopoles. They showed that they have a structure consistent with their interpretation as heavy gauge particles of an overall symmetry group, $G^\vee$, dual to $G$, providing further circumstantial evidence for the some form of Montonen-Olive duality. 

\bigskip
\bigskip
\bigskip
\centerline{\sc Toda Theories}
\bigskip
\nobreak
David Olive continued his studies of solutions to gauge theories with symmetry broken by a Higgs field in the adjoint representation, working with  Nikos Ganoulis and Peter Goddard (65), investigating static stable solutions to the field equations in the BPS limit that are spherically symmetric in a suitable sense. Andrej Leznov and Mikhail Saveliev (1980) had shown that they could be constructed from solutions to a spatial version of the so-called Toda molecule equations,   
\be
 {d^2\theta_i\over dr^2}= \exp\left(\sum_{j=1}^RK_{ij}\theta_j\right), \label{t1}
\ee
where $K_{ij} = 2\balpha_i\cdot\balpha_j/\balpha_j^2$ is the Cartan matrix of $G$, $\balpha_i, 1\leq i\leq R$, being a basis of simple roots. The equations  [\ref{t1}]
are integrable and Ganoulis, Goddard and Olive showed how to determine the constants of integration so as to ensure regularity at the origin, thus obtaining solutions for any simple group $G$ in terms of its root system. This work marked the beginning of David's fascination with Toda systems, which he kept coming back to, resulting in a series of contributions over the following fifteen years. 

 Toda's work (1967) was motivated by the numerical studies of  Fermi, Pasta and Ulam (1955) of the behaviour of a system of particles joined by springs exerting forces nonlinear in their extensions. Toda made the important observation that if the polynomial forces considered by Fermi {\it et al.} were replaced by exponential ones, a one-dimensional lattice comprising an infinite number of such particles could be solved analytically instead of numerically. It was found that this integrability extended to finite systems of particles, labelled by the simple roots of finite-dimensional Lie algebras, as in [\ref{t1}], replacing $d^2\theta_i/dr^2$ by $-d^2\theta_i /dt^2$ (Kostant, 1979), and also  to corresponding systems of nonlinear relativistic wave equations, {\it i.e.} Toda field theories,  replacing $d^2\theta_i/dr^2$ by $d^2\theta_i/dr^2 -d^2\theta_i /dt^2$ in [\ref{t1}]. There are also infinite lattice systems similarly related to the simple root systems of Kac-Moody algebras (Mikhailov, Olshanetsky and Perelomov 1981). 

With Neil Turok (70), David showed how the symmetries of the Dynkin diagram of a simple Lie algebra led to new integrable systems of equations, which are reductions of the Toda molecule equations associated with the algebra, obtained by identifying variables related by the symmetry. Then, turning to the Toda lattice field theories associated with  Kac-Moody algebras, termed affine Toda field theory,
Olive and Turok  systematically constructed an infinite number of commuting local conserved quantities, which could be used as Hamiltonians to define evolution in associated times (75, 86).  

David returned to the study of Toda theories  in the 1990s, following work by others on exact S-matrices for two-dimensional integrable quantum field theories, starting with the sine-Gordon quantum field theory, which is actually the simplest of the Toda field theories, being based on  the group $SU(2)$. A key feature of the sine-Gordon theory is that classically it possesses soliton solutions, including soliton-anti-soliton bound states, known as breather solutions. These persist in the quantum field theory, becoming a finite discrete set of states whose number depends on the coupling constant.
David's approach was to seek to use general Lie algebraic techniques, which he was familiar with from his work on magnetic monopoles and on conformal field theory, to provide unified and deeper understandings of mass spectra and couplings in affine Toda field theories (100, 101), in particular the rule describing three-point couplings discovered by Patrick Dorey (1991). Given David's background, it was natural for him to apply these insights to the conjectured  $S$-matrix of the quantum field theory.

Typically the soliton solutions are complex rather than real, and so might seem uninteresting physically. However, with Turok and David's student, Jonathan Underwood (102,  103), he showed that the solutions have real energy and momentum, equal to the sum of contributions of the individual solitons. They further showed that the number of species of soliton equalled the rank of the Lie group with which it was associated and their masses were given in terms of its algebraic structure, and used a vertex operator formalism for constructing soliton solutions (105). With Saveliev and Underwood (104), and also with his students Andreas Fring, Peter Johnson, and Marco Kneipp (108, 110), David used vertex operators to study solitons and their scattering further, calculating time delays and showing that the process should be interpreted as transmission slowed down by attractive forces rather than reflection. 
No doubt one of the attractions of the study of affine Toda field theories for David was that aspects of them  bore some resemblance to Yang-Mills-Higgs theories in four dimensions, with their monopole solutions analogous to the solitons of the Toda theories, and the mass spectrum of the solitons being equal to the mass spectrum of the dual theory based on the dual affine Lie algebra,  but with the inverse coupling constant, recalling Montonen-Olive duality in gauge theories.

\bigskip
\bigskip
\bigskip
\centerline{\sc Algebras, Lattices and Strings}
\bigskip
\nobreak
Having become aware of the profound connections that had been discovered between the vertex operators developed in dual models and string theory, on the one hand, and representations of Kac-Moody algebras, on the other (Frenkel \& Kac 1980, Segal 1981), about 1980 David began, with Peter Goddard, to explore the possible role of these algebras in physics, and  in string theory in particular.
A further motivation was a proof by Graeme Segal (1980) of the Jacobi identity using a fermion-boson equivalence, the isomorphism between the space of states for a single boson field defined on a circle and that for a pair of fermion fields. The role of integral lattices ({\it i.e.}~ones such that the scalar product between any two points is an integer) in these constructions provided links with concepts David had introduced in the theory of magnetic monopoles. 

In Charlottesville in 1983, Goddard and Olive wrote an account synthesizing some of what they had learned and found out, under the title, {\it Algebras, Lattices and Strings} (72). They used dual model vertex operators to provide a unified construction of finite dimensional Lie algebras, affine Kac-Moody algebras, Lorentzian algebras and fermionic extensions of these algebras. Given an integral lattice, $\Lambda$, they associated to each point, $r\in\Lambda$, of squared length 2, the operator  $e_r= c_rA_r$, where  $A_r$ is the contour integral about the origin of the dual model (or string) vertex operator, $V(r,z)$, for emitting a `tachyon', 
\be\label{a1}
A_r
={1\over 2\pi i}\oint  V(r,z)\,{dz\over z}\,,
\ee
and $c_r$ is a function of momentum such that $c_rc_s= (-1)^{r\cdot s}c_sc_r$. They considered the Lie algebra, $\g_\Lambda$, generated by these $e_r$ under commutation.  If $\Lambda$  contains points of squared length 1,
$\g_\Lambda$ may be extended into a superalgebra by including operators associated with those points, which quite naturally satisfy anti-commutation relations.

If the scalar product on  $\Lambda$ is positive definite, $\g_\Lambda$ is a compact finite-dimensional Lie algebra, with a basis comprising the $A_r$, with $r\in\Lambda$ and $r^2=2,$ together with the momenta.  If the scalar product is positive semi-definite,  $\g_\Lambda$ is an affine Kac-Moody algebra, which includes the contour integrals of 
 vertex operators for emitting `photons', associated with points $k\in\Lambda$ with $k^2=0,$ as well as momenta. If the scalar product is indefinite, $\g_\Lambda$ involves the vertex operators associated with other dual model states. 
 
In the context of string theory, if $\Lambda$ were a lattice of momenta of physical states, discrete because the corresponding spatial dimensions were compactified,  $\g_\Lambda$ would map the physical state space into itself, {\it i.e.} the physical states would fall into representations of  $\g_\Lambda$. Motivated in a general way by the ideas of electromagnetic duality that he had pioneered in gauge theories, where generalized electric and magnetic charges correspond to points of dual lattices, David formed the  intuition that the cases where $\Lambda$ is self-dual  might be particularly interesting. Further, to concentrate on the bosonic case, Goddard and Olive considered the case where $\Lambda$ is even, {\it i.e.} all the squared lengths of all  its points are even, as well as self-dual. 

The requirement of  $\Lambda$ being both even and self-dual is quite restrictive; for $ \Lambda\subset \Rop^{m,n}$, it is necessary that $m-n$ be a multiple of 8. Thus, in the Euclidean case, the smallest possibilities for the dimension of $\Lambda$ are $8, 18, 24$, and these are the only dimensions relevant to bosonic string theory, for which the space-time dimension is bounded by 26. Goddard and Olive noted that the only such lattice with $\dim\Lambda=8$ is the root lattice of $\Er_8$, for 
 $\dim\Lambda=16$ there are two possibilities: the root lattice of $\Er_8\times\Er_8$ and a sublattice of the weight lattice of the Lie algebra of $\hbox{SO}(32)$ (corresponding to the weight lattice of the group $\hbox{Spin}(32)/\Zop_2$), while for  $\dim\Lambda=24$ there are 24 possibilities.  

Applied to the quantum motion of a relativistic string, moving in a space some of whose dimensions are compactified to form a torus by identifying points of space related by displacements forming an even self-dual lattice (the length scale being set by the characteristic length of the string), $\g_\Lambda$ generates a gauge symmetry. For a closed string, the symmetry would be   $\g_\Lambda\oplus \g_\Lambda$. To avoid this, Gross, Harvey, Martinec and Rohm (1985) found it was necessary to treat left and right moving waves on the closed string differently, so that, somewhat bizarrely, only those moving in one direction had motion in these compactified directions. The model they constructed, the heterotic string, enabled them to realize in string theory either of the gauge groups, $\Er_8\times\Er_8$ or  $\hbox{Spin}(32)/\Zop_2$, for which Green and  Schwarz (1984) had found anomaly cancellations for supersymmetric gauge theories coupled to gravity.

Rather than pursue these applications of Kac-Moody algebras, David returned to investigations of the equivalence of  descriptions of two-dimensional quantum field theories in terms of either boson fields or fermion fields. Edward Witten (1984) had extended the known results on the equivalence in two dimensions between a single boson field and a pair of fermion fields to an equivalence between certain nonlinear boson theories and  certain free fermion theories. Specifically, he demonstrated an equivalence between a nonlinear $\sigma$ model, with a field $g$ taking values in the group  $\hbox{SO}(N)$, and a theory with $N$ free fermion fields $\psi^i,1\leq i\leq N.$ 

To obtain suitable equations of motion for $g$, Witten needed to include a Wess-Zumino term in the $\sigma$-model Lagrangian, defining what is now known as the Wess-Zumino-Witten (WZW) theory, so that these then become 
$\partial_+(g^{-1}\partial_-g)=\partial_-(\partial_+gg^{-1})=0$, 
where $\partial_\pm$ denote the partial derivatives with respect to 
$x^\pm=x\pm t$, implying that $\partial_+gg^{-1} $ and
$g^{-1}\partial_-g$ depend only on $x^+$ and $x^-$, respectively. Imposing periodic  boundary conditions, identifying $x$ with $x+L$, and expanding  $\partial_+gg^{-1} $ in powers of $z=e^{2\pi ix^+/L}$ and a basis, $\{t^a\}$, for the Lie algebra of $\hbox{SO}(N)$, the canonical commutation relations imply that, with  suitable normalizations, its components, $T^a_n$, satisfy the Kac-Moody algebra,
\be \label{a2}
[T^a_m,T^b_n]=i{f^{ab}}_cT^c_{m+n}+km\delta^{ab}\delta_{m,-n},
\ee
where $k$ is an integer determined by the coefficient of the Wess-Zumino term, and ${f^{ab}}_c $ are structure constants for $\hbox{SO}(N)$. Witten observed that a representation same algebra was provided by
\be \label{a3}
T^a_n\rightarrow {i\over 2}\sum_r\psi^i_r M^a_{ij}\psi^j_{n-r}
\ee
where the $M^a$ are the representation matrices for the defining representation of the Lie algebra of $\hbox{SO}(N)$ and $\psi^i_r$ are the modes of the free fermion fields labelled by half odd integers $r$ (for convenience taken to be odd under $x \rightarrow x+L$). The Kac-Moody algebra [\ref{a2}] with $k=1$ follows from the canonical anti-commutation relations, $\{\psi^i_r,\psi^j_s\}=\delta_{r,-s}\delta^{ij}$. Witten observed that [\ref{a2}] has only a few positive energy unitary representations for $k=1$, so that the isomorphism of the Kac-Moody algebras in the two theories effectively guarantees their equivalence in this case. 

The requirements of positive energy, {\it i.e.}~that the spectrum of $L_0$ be non-negative, and unitary require $k$ in [\ref{a2}] to be a positive integer. (In the case of the affine Kac-Moody algebra $\hat\g$ associated with a general compact simple Lie algebra $\g$, the condition is that $\x=2k/\varphi^2$ should be a positive integer, called the level of the representation.) For $\hbox{SO}(N)$ WZW theories with $k>1$, Witten proposed taking $k$ copies of the $N$ fermion theory, but, as David surmised might be the case, the equivalence is then no longer exact: there is more in the fermion theory than that WZW model.

David thought that exact equivalence would entail not only the isomorphism of the Kac-Moody algebras possessed by the two theories, but also this should extend to an equivalence of the energy-momentum tensors in the boson and fermion theories. Working in the general context of the WZW theory for a simple Lie group, $G$, with an $N$-dimensional real representation, $M$ in [\ref{a3}], Goddard and Olive (74) sought to determine whether the energy-momentum tensors of the WZW theory and the $N$ fermion theory were necessarily identical, and, if not, what conditions would ensure that they were equal. 

The conformal symmetry of each of these theories means that the modes of the energy momentum tensor satisfy the Virasoro algebra,
\be \label{a4}
[L_m,L_n]=(m-n)L_{m+n}+{c\over 12}m(m^2-1)\delta_{m,-n}\,,
\ee
with $L_n=\L^\g_n, c=c^\g$, where
\be \label{a5}
c^\g ={2k\dim\g\over 2k+Q^\g} ={\x\dim\g\over \x+h^\g}\,,
\ee

for the WZW theory, and $L_n=L^\psi_n, c=c^\psi=\half N$ for the fermion theory, where 
\be \label{a6}
\L^\g_n={1\over 2k+Q_\varphi} \sum_m\nox T^a_{-m} T^a_{n+m} \nox
  \,,\qquad L^\psi_n= {1\over 2} \sum_r :rb_{-r}^jb_{n+r}^j:\,.
\ee
[Here $\nox$ and $:$ denote normal ordering with respect to the modes $T^a_m $ and $b^i_r$, respectively, and $Q^\g$ is the quadratic Casimir operator for the adjoint representation of $\g$, and $h^\g=Q^\g/\varphi^2$, the dual Coxeter of $\g$.] For the affine algebra $\hat g$
to be the same for the two theories, we need $k=\half\kappa$, where $M^a_{ij}\,M^b_{ij}=\kappa\,\delta^{ab}$.

If the two energy-momentum tensors are equal, it is necessary that $c^\g=c^\psi$.
Goddard and Olive found that  this is typically not  the case, and, further, that the difference $K_n=L^\psi_n-\L^\g_n$ itself satisfies a Virasoro algebra, commuting with $\L^\g_m$,
{\it i.e.}~[\ref{a4}] holds for $L_n=K_n$ and $c=c^\psi-c^\g$. The requirement that the Virasoro algebra representations provided by $L^\psi_n$ and $\L^\g_n$ are unitarity and positive energy implies that  $K_n$ has the same property and so that $c^\g\leq c^\psi $. If $c^\g= c^\psi $,  $K_n$ vanishes because the Virasoro algebra does not have any non-vanishing positive energy unitary representations with $c=0$. Thus  $c^\g= c^\psi$ provides a condition for the identity of the energy-momentum  tensors and the equivalence of the theories that is both necessary and sufficient.

The cases where $c^K=c^\psi-c^\g>0$ also proved to be of great interest. The positive energy unitary representations of the Virasoro algebra [\ref{a4}] are labelled by two parameters, $c$ and the lowest eigenvalue of $L_0$, $h$, say. Some months before Goddard and Olive had begun their study of WZW models, 
Friedan, Qiu and Shenker (FQS) had shown that, for such representations, either $c \geq 1$ or $c$ belongs to an infinite discrete sequence,  
\be\label{a7}
c = 1 - {6 \over {(m+2)(m+3)}}, \qquad 
m = 0, \, 1, \,\dots \,,
\ee
for each value of which there are $  {1\over 2} (m+1)(m+2)$ possible values of $h$ (Friedan et al. 1984). At that time, it was known whether any of these discrete series representations exist, apart from $c=0$, the trivial representation, and $c=  {1\over 2} $, which corresponds to a single fermion field; the rest, with $c=  {7\over10},\, {4\over 5}, \, {6\over 7}, \,\ldots\,,$ were unknown. In most cases either $c^K=0$ or $c^K\geq 1$, but Goddard and Olive found that taking $M^a$ to be the 7-dimensional representation of $G=\hbox{SO}(3)$ gave the  $c^K=  {7\over10}$ representation of the Virasoro algebra, with all the associated 6 values of $h$.

It turns out that the other values of $c$ in the discrete series [\ref{a7}], those with $m\geq 4$, can not be obtained in this way and a more general approach is needed. This was found by Goddard, Kent and Olive (GKO) who changed the context from the free fermion representation of the affine algebra associated with $\hbox{so}(N)$ to that a positive energy unitary representation of the affine algebra, $\hat\g$, associated with compact simple Lie algebra, $\g'$, and a subalgebra $\g\subset\g'$. The representation of $\hat\g'$ provides a representation of $\hat\g$, whose level $\x$ is determined by the level $\y$ of the representation of $\hat\g'$. (We obtain the previous construction by taking $\g'=\hbox{so}(n)$.) We may define a Virasoro representation $\L^{\g'}_n$ associated to the representation of $\hat\g'$ in the same way the Virasoro representation $\L^\g_n$ is associated to the representation of $\hat\g$ by [\ref{a6}]. Now  $K_n=\L^{\g'}_n - \L^\g_n$ commutes with the whole of $\hat\g$ and so with $\L^\g_n$. It follows that $K_n$ provides a representation of the Virasoro algebra with central charge,
\be \label{a8}
c^K=c^{\g'} -c^\g = {\y\dim\g'\over \y+h^{\g'}}-  {\x\dim\g\over \x+h^\g}
\,.
\ee
This construction, which has become known as the coset (or GKO) construction (76), extends straightforwardly to cases where $\g'$ or $\g$ are not simple (in which cases the various the level is defined for each summand).
All of the values in the series [\ref{a7}], and all the associated values of $h$, can be obtained in this way, {\it e.g.}~by taking $\g'=\hbox{sp}(m+1), \x=1, \g=\hbox{sp}(m)\oplus \hbox{sp}(1)$, for both of which the level is also 1 (81). 

Given a conformal field theory 
(CFT) associated with the affine Kac-Moody algebra $\hat\g'$, and $\g\subset\g'$, the coset construction defines a CFT associated with $\g'/\g$, on the subspace of states annihilated by the positive modes of $\hat\g$, with Virasoro algebra $K_n$. If $c^K=0$, this is trivial, and the CFT associated with $\hat\g$ is said to be conformally embedded in that associated with $\hat\g'$. The coset construction has provided one if the principal ways of constructing interesting examples of CFTs in applications. 

David returned to the investigation of the condition under which the energy-momentum tensors of a WZW theory for the group $G$,  with an $N$-dimensional real representation, $M$, and an $N$ fermion theory are equal, from which the coset construction arose as a by-product, and, with Peter Goddard and Werner Nahm, he found an elegant geometric interpretation, which enabled the  classification of the cases where it is satisfied (78).  The condition for the vanishing of $K_n=L^\psi_n-\L^\g_n$ is
\be \label{a9}
M^a_{ij} M^a_{kl}+ M^a_{ik} M^a_{lj}+ M^a_{il} M^a_{jk}=0\,.
\ee
They observed that this is the condition that $\g$ can be extended by an $N$-dimensional space, $\s$, to give  a Lie algebra,  $\g'=\g\oplus\s$, which is a symmetric space, {\it i.e.}~$[\g,\g]\subset\g,\, [\g,\s]\subset\s, \,[\s,\s]\subset\g,$ with the action of $\g$ on $\s$ providing the representation $M$. Using $\{s_i\}$ to denote an orthonormal basis for $\s$, for  $\g'=\g\oplus\s$ to be a symmetric space, the communication relations must have the form, 
\be\label{a10}
[t^a,t^b]=i {f^{ab}}_ct^c\,,\qquad
[t^a,s_i]=i M^a_{ij}s_j\,,\qquad
[s_i,s_j]=i M^a_{ij}t^a\,.
\ee
Given that $\g$ is a Lie algebra with representation $M$ 
the condition for these to be consistent is the Jacobi identity for $s_i,s_j,s_k$, which is just [\ref{a9}].
Symmetric spaces have been classified and so provide a list of the instances in which $\L^\g_n=L^\psi_n$.

\bigskip\vskip6pt

\centerline{\large\sc Acknowledgements}
\small
In addition to his account of his work on string theory (44), David Olive left detailed notes on his early life and much of his scientific career, which have been immensely valuable to us.
We are very grateful to many of David Olive's friends and colleagues for much information and helpful correspondence and, most particularly, to David's widow, Jenny Olive, and younger daughter, Rosalind Shufflebotham, who also read carefully drafts of the biography, and provided the frontispiece and other photographs. Jenny died on 1 September 2018.

\vfil\eject
\phantom{x}\bigskip
\centerline{\large\sc References to other authors}
\def\hi{\hangindent=40pt}
\def\NC{\it Nuovo Cim. \rm}
\def\NP{\it Nucl. Phys. \rm}
\def\PR{\it Phys. Rev. \rm}
\def\PRL{\it Phys. Rev.  Lett. \rm}
\def\PL{\it Phys. Lett. \rm}
\def\JMP{\it J. Math. Phys. \rm}
\def\LMP{\it Lett. Math. Phys. \rm}
\def\CMP{\it Commun. Math. Phys. \rm}
\def\PRt{\it Phys. Rept. \rm}
\def\RPP{\it Rep. Prog. Phys. \rm}
\def\RNC{\it Rev. Nuovo Cim. \rm}

\bigskip\bigskip

\hi 
Bogomol'nyi, E. B. 1976 The stability of classical solutions. {\it Sov. J. Nucl. Phys.} {\bf 24}, 389--394.

\hi
Chew, G. F. 1966 The Analytic S Matrix. New York, 1966: Benjamin.

\hi
Coleman, S. 1975 Quantum sine-Gordon equation as the massive Thirring model. \PR {\bf D11}, 2088--2097.

%

\hi
Cutkosky, R. E.  1960 Singularities and discontinuities of Feynman amplitudes. \JMP {\bf 1}, 429--433.

\hi
Dirac, P. A. M. 1931 Quantised singularities in the electromagnetic field. {\it Proc. Roy. Soc.} {\bf A33}, 60--72.

\hi
Dorey, P. 1991 Root systems and purely elastic $S$-matrices. \NP {\bf B358}, 654--676.

\hi
Fermi, E., Pasta,  J. \& Ulam S. 1955 Studies of nonlinear problems. I. Los Alamos report LA-1940, published later in E. Segr\'e (ed.) 1965 {\it Collected Papers of Enrico Fermi} , University of Chicago Press.

\hi 
Frenkel, I. B. \& Kac, V. G. 1980 Basic representations of affine Lie algebras and dual resonance models. {\it Invent. Math.} {\bf 62}, 23--66.

\hi
Friedan, D., Qiu, Z. \& Shenker, S. 1984 Conformal invariance, unitarity, and critical exponents in two dimensions. \PRL {\bf 52}, 1575--1578.

\hi
Goddard, P., Goldstone, J.,  Rebbi, C. \& Thorn, C. B. 1973 Quantum dynamics of a massless
relativistic string. \NP {\bf B56}, 109--135.

\hi
Green, M. B. \& Schwarz, J. H. 1984
Anomaly cancellations in supersymmetric $D = 10$ gauge theory and superstring theory. \PL {\bf 149B}, 117--122.

\hi
Gross, D. J., Harvey, J. A., Martinec, E. \& Rohm, R. 1985  Heterotic string theory (I) The free heterotic string. \NP {\bf B256}, 253--284.

\hi 
Gunson, J. 1965 Unitarity and on-mass-shell analyticity as a basis for S-matrix theories. I, II, III. \JMP {\bf 6}, 827--844; 845--851; 852--858.


\hi
Iagolnitzer, D. 1981 Analyticity properties of the S-matrix: historical survey and recent results in S-matrix theory and axiomatic field theory. {\it Acta Phys. Austriaca, Suppl.,} {\bf 23}, 235--328.

\hi
Kac, V. G. 1967 Simple graded algebras of finite growth. {\it Funct. Anal. Appl.} {\bf 1}, 328--329.

\hi
Kostant, B. 1979 The solution to a generalized Toda lattice and representation theory. {\it Advances in Mathematics} {\bf 34}, 195--338.

\hi
Landau, L. D. 1959
On analytic properties of vertex parts in quantum field theory. \NP {\bf 13}, 181--192.

\hi
Langlands, R. P. 1970 Problems in the theory of automorphic forms In: {\it Lectures in Modern Analysis and Applications III} (ed. C. T. Taam) {\it Springer Lecture Notes in Mathematics} {\bf 170}, pp. 18--61.

\hi
Leznov, A. N. \& Saveliev, M. V. 1980 Representation theory and integration of nonlinear spherically symmetric equations to gauge theories. \CMP {\bf 74}, 111--118.

\hi
Lovelace, C. 1971 Pomeron form factors and dual Regge cuts. \PL  {\bf 34B} 500--506.

\hi
Manton, N. S. 1977 The force between 't Hooft-Polyakov monopoles. \NP {\bf B126}, 525--541.

\hi
Mandelstam, S. 1959  Analytic properties of transition amplitudes in perturbation theory. \PR {\bf 115} 1741--1751; 
Construction of the Perturbation Series for Transition Amplitudes from their Analyticity and Unitarity Properties. \PR {\bf 115} 1752--1762.

\hi
Mandelstam, S. 1973 Manifestly dual formulation of the Ramond model. Phys. Lett. {\bf 46B} 447--451.

\hi
Mandelstam, S. 1975 Soliton operators for the quantized sine-Gordon equation. \PR {\bf D11},  3026--3030. 

\hi
Moody, R. V. 1967 Lie algebras associated with generalized Cartan matrices. {\it Bull. Amer. Math. Soc.} {\bf 73}, 217--221.

\hi
Mikhailov, A. V., Olshanetsky, M. A. \& Perelomov, A. M. 1981 Two dimensional generalized Toda lattice.  \CMP {\bf 79}, 473--488.

\hi
Monastyrski\u i, M. I.  \& Perelomov, A. M. 1975 Concerning the existence of monopoles in gauge field theories. {\it ZhETF Pis. Red.} {\bf 21},  94--96. 

\hi
Nambu, Y. 1969 Quark model and the factorization of the Veneziano amplitude. In: {\it Proceedings
of the International Conference on Symmetries and Quark Models held at Wayne
State University, June 18--20, 1969} (ed. R. Chand) pp. 258--277. New York: Gordon and Breach.

\hi
Neveu, A. \& Schwarz, J. H. 1971 Quark model of dual pions. \PR {\bf D4}, 1109--1111.

\hi
Nielsen, H. B.  1969 An almost Physical Interpretation of the integrand of the $n$-point
Veneziano model (unpublished).

\hi
Osborn, H. 1979
Topological charges for $N=4$ supersymmetric gauge theories and monopoles of spin $1$. \PL {\bf 83B}, 321--326.

\hi
Polyakov,  A.  M.  1974 Particle spectrum in quantum field theory. {\it JETP Lett.} {\bf 20}, 194--195.

\hi
Prasad,  M.  K.  \& Sommerfield,  C. M. 1975 An exact classical solution for the 't Hooft monopole and the Julia-Zee dyon. \PRL  {\bf 35}, 760--762.

\hi
Ramond, P.  1971 Dual theory for free fermions. \PR  {\bf D3}, 2415--2418.

\hi
Salam, A. \& Komar, A. 1960 Renormalization problem for vector meson theories. \NP {\bf 21}, 624--630.

\hi 
Sato, M. 1975 Recent development in hyperfunction theory and its applications to physics (microlocal analysis of S-matrices and related quantities) {\it Lect. Notes Math.} {\bf 449}, 13--29. Berlin: Springer-Verlag.

\hi
Schwarz, J. H. 1971 Dual quark-gluon model of hadrons. \PL {\bf 37B}, 315--319.

\hi
Schwarz, J. H. \& Wu, C. C. 1974 Functions occurring in dual fermion amplitudes. \NP  {\bf B73} 77--92.

\hi
Segal, G. 1980 Jacobi's identity and an isomorphism between a symmetric algebra and an exterior algebra (unpublished).

\hi
Segal, G. 1981 Unitary representations of some infinite dimensional groups. \CMP {\bf 80}, 301--342.

\hi
Seiberg, N. \& Witten, E. 1994a Electromagnetic duality, monopole condensation, and confinement in $N = 2$ supersymmetric
Yang-Mills theory. \NP {\bf B426}, 19--52; Erratum {\bf B430}, 485--486.

\hi
Seiberg, N. \& Witten, E. 1994b Monopoles, duality and
chiral symmetry breaking in N = 2 supersymmetric QCD.  \NP {\bf B431}, 484--550. 

\hi
Sen, A 1994. Dyon-monopole bound states, self-dual harmonic
forms on the multi-monopole moduli space, and SL(2, $\Zop$) invariance in string
theory. \PL {\bf B298}, 217--221.

\hi
Skyrme, T. H. R.  1961 Particle states of a quantized meson field. {\it Proc. Roy. Soc.} {\bf A262}, 237--245. 

\hi
Stapp, H. P.  1962 Derivation of the CPT theorem and the connection between spin and statistics from postulates of the S-Matrix theory. \PR {\bf 125}, 2139--2162. 

\hi
Streater, R. \& Wightman, A. 1964 PCT, Spin and Statistics, and All That. Princeton: Princeton University Press. 

\hi
Susskind, L.  1970 Dual-symmetric theory of hadrons I. \NC {\bf 59A}, 457--496.

\hi
't Hooft, G. 1974 Magnetic monopoles in unified gauge theories. \NP {\bf B79}, 276--281.

\hi
Thorn, C. B. 1971 Embryonic dual model for pions and fermions \PR {\bf D4}, 1112--1116.

\hi Toda, M. 1967 Vibration of a chain with nonlinear interaction. {\it J. Phys. Soc. Japan} {\bf 22}, 431--436.

\hi
Tyupkin, Yu. S., Fateev, V. A. \& Shvarts, A. S. 1975
Existence of heavy particles in gauge field theories. {\it ZhETF Pis. Red.} {\bf 21},  91--93. 

\hi
Veneziano, G. 1968 Construction of a crossing-symmetric, Regge-behaved amplitude for
linearly rising trajectories. \NC {\bf 57A}, 190--197.

\hi
Witten, E. 1984 Non-abelian bosonization in two dimensions. \CMP {\bf 95}, 455-472.

\bigskip\bigskip\bigskip
\centerline{\sc\large Bibliography}

\bigskip

\begin{tabular}{r r p{15cm}}
(1)&1962&(with J. C. Taylor) Singularities of scattering amplitudes at isolated real points. \NC {\bf 24}, 814--822.\\
(2)&1962&Unitarity and evaluation of discontinuities. \NC {\bf 26}, 73--102.\\
(3)&1963&On analytic continuation of scattering amplitude through a three-particle cut. \NC {\bf 28}, 1318--1336.\\
(4)&1963&Unitarity and evaluation of discontinuities -- II. \NC {\bf 29}, 326--335.\\
(5)&1964&Exploration of $S$-matrix theory. \PR {\bf 135},  B745--B760.\\
(6)&1965&Unitarity and evaluation of discontinuities -- III. \NC {\bf 37}, 1423--1445.\\
(7)&1966&(with P. V. Landshoff \& J. C. Polkinghorne) Hierarchical principle in perturbation theory. \NC {\bf A43},  444--453.\\
\end{tabular}

\begin{tabular}{r r p{15cm}}
(8)&1966&(with E. Y. C. Lu) Spin and statistics in $S$-matrix theory. \NC {\bf  45A},  205--218.\\
(9)&1966&(with P. V. Landshoff \& J. C. Polkinghorne) Icecream-cone singularity in $S$-matrix theory. \JMP {\bf 7}, 1600--1606.\\
(10)&1966&(with P. V. Landshoff \& J. C. Polkinghorne) Normal thresholds in subenergy variables \JMP {\bf 7},  1593--1599.\\
(11)&1966&(with P. V. Landshoff) Extraction of singularities from $S$-matrix. \JMP  {\bf 7},  1464--1477.\\
(12)&1966&(with R. J. Eden, P. V. Landshoff \& J. C. Polkinghorne) {\it The Analytic $S$-Matrix}. Cambridge, UK: Cambridge University Press.\\
(13)&1968&(with J. C. Polkinghorne) Properties of Mandelstam cuts. \PR {\bf 171}, 1475--1481.\\
(14)&1969&(with R. E. Cutkosky, P. V. Landshoff \& J. C. Polkinghorne) A non-analytic $S$-matrix. \NP {\bf B12},  281--300.\\
(15)&1969&(with M. J. W. Bloxham, D. I. Olive \& J. C. Polkinghorne) $S$-matrix singularity structure in the physical region. I. Properties of multiple integrals.
\JMP {\bf 10}, 494--502.\\
(16)&1969&(with M. J. W. Bloxham, D. I. Olive \& J. C. Polkinghorne) $S$-matrix singularity structure in the physical region. II. Unitarity integrals.
\JMP {\bf 10},  545--552.\\
(17)&1969&(with M. J. W. Bloxham, D. I. Olive \& J. C. Polkinghorne) $S$-matrix singularity structure in the physical region. III. General discussion of simple Landau singularities. \JMP {\bf 10}, 553--561.\\
(18)&1969&(with D. K. Campbell \& W. J. Zakrzewski) Veneziano amplitudes for Reggeons and spinning particles. \NP {\bf  B14}, 319--329.\\
(19)&1969&(with W. J. Zakrzewski) A Veneziano amplitude for four spinning pions. \PL {\bf B30}, 650--652.\\
(20)&1970&(with W. J. Zakrzewski) A possible Veneziano amplitude for many pions. \NP {\bf B21}, 303--320.\\
(21)&1970&The connection between Toller and Froissart-Gribov signatured amplitudes. \NP {\bf B15}, 617--627.\\
(22)&1970&(with V. Alessandrini, D. Amati \& M. Le Bellac) Duality and gauge properties of twisted propagators in multi-Veneziano theory.
\PL {\bf B32}, 285--290.\\
(23)&1970&(with D. Amati \& M. Le Bellac) Twisting-invariant factorization of multiparticle dual amplitude. \NC {\bf A66}, 815--830.\\
(24)&1970&(with D. Amati \& M. Le Bellac) The twisting operator in multi-Veneziano theory. \NC {\bf A66}, 831--844.\\
(25)&1971&Operator vertices and propagators in dual theories. \NC {\bf A3}, 399--411.\\
(26)&1971&(with V. Alessandrini, D. Amati \& M. Le Bellac) The operator approach to dual multiparticle theory. \PRt {\bf 1C}, 269--345.\\
(27)&1971&The dual approach to the strong interaction $S$-matrix. In {\it Proceedings of the Symposium on Basic Questions in Elementary Particle.} pp. 140--147. M\"unchen, Germany: Max-Planck-lnstitut f\"ur Physik und Astrophysik. \\ 
(28)&1971&(with E. Corrigan) Fermion meson vertices in dual theories. \NC {\bf A11}, 749--773.\\
(29)&1972&Clarification of the rubber string picture. In: {\it Proceedings of 16th International Conference on High-energy Physics, Batavia, IL} (ed. J. D. Jackson \& A. Roberts), vol. 1, pp.  472-474. Batavia, IL, USA: National Accelerator Laboratory. \\
\end{tabular}

\begin{tabular}{r r p{15cm}}
(30)&1973&(with L. Brink \& D. I. Olive) The physical state projection operator in dual resonance models for the critical dimension
of space-time. \NP {\bf B56}, 253--265.\\
(31)&1973&(with L. Brink) Recalculation of the unitary single planar dual loop in the critical dimension of spacetime. \NP {\bf B58}, 237--253.\\
(32)&1973&(with L. Brink, C. Rebbi \& J. Scherk) The missing gauge conditions for the dual fermion emission vertex and their consequences.
\PL {\bf B45}, 379--383.\\
(33)&1973&(with L. Brink \& J. Scherk) The gauge properties of the dual model pomeron-reggeon vertex -- their derivation and their
consequences. \NP {\bf B61}, 173--198.\\
(34)&1973&(with J. Scherk) No ghost theorem for the pomeron sector of the dual model. \PL {\bf B44}, 296--300.\\
(35)&1974&(with J. Scherk) Towards satisfactory scattering amplitudes for dual fermions.
\NP {\bf B64}, 334--348.\\
(36)&1973&(with E. Corrigan, P. Goddard \& R. A. Smith) Evaluation of the scattering amplitude for four dual fermions.
\NP {\bf B67}, 477--491.\\
(37)&1975&Unitarity and discontinuity formulas. {\it Lect. Notes Math.} {\bf 449}, 133--142.
Berlin: Springer-Verlag.\\
(38)&1975&Dual Models. In: {\it Proceedings of 17th International Conference on High-energy Physics, London} (ed. J.R. Smith), pp.  269-280. Chilton, UK: Rutherford Laboratory. \\
(39)&1975&(with D. Bruce \& E. Corrigan) Group theoretical calculation of traces and determinants occurring in dual theories.
\NP {\bf B95}, 427--433.\\
(40)&1976&(with E. Corrigan, D. B. Fairlie \& J. Nuyts) Magnetic monopoles in SU(3) gauge theories. \NP {\bf B106}, 475--492.\\
(41)&1976&Some topics in dual theory. {\it Acta Universitatis Wratislaviensis} {\bf 368}, 293--310.\\ 
(42)&1976&(with E. Corrigan) Color and magnetic monopoles. \NP {\bf B110}, 237--247.\\
(43)&1976&Angular momentum, magnetic monopoles and gauge theories. \NP {\bf B113}, 413--420.\\
(44)&1976&(with F. Gliozzi \& J. Scherk) Supergravity and the spinor dual model. \PL {\bf B65}, 282--286.\\
(45)&1977&(with F. Gliozzi \& J. Scherk) Supersymmetry, supergravity theories and the dual spinor model.
\NP {\bf B122}, 253--290.\\
(46)&1977&(with P. Goddard \& J. Nuyts) Gauge theories and magnetic charge. \NP {\bf B125}, 1--28.\\
(47)&1977&(with C. Montonen) Magnetic monopoles as gauge particles? \PL {\bf B72}, 117--120.\\
(48)&1977&Progress in analytic $S$-matrix theory. {\it Publ. Res. Inst. Math. Sci. Kyoto} {\bf 12}, 347--349.\\
(49)&1978&(with P. Goddard) Magnetic monopoles in gauge field theories. \RPP {\bf 41}, 1357--1437.\\
(50)&1978&(with E. Witten) Supersymmetry algebras that include topological charges. \PL {\bf B78}, 97--101.\\
(51)&1979&Supersymmetric solitons. {\it Czech. J. Phys.} {\bf B29},  73--80.\\
(52)&1979&Magnetic monopoles. \PRt {\bf 49},  165--172.\\
(53)&1979&The electric and magnetic charges as extra components of four momentum. \NP {\bf B153}, 1--12.\\
(54)&1979&(with S. Sciuto \& R. J. Crewther) Instantons in field theory. \RNC {\bf 2}, 1--117.\\
\end{tabular}

\begin{tabular}{r r p{15cm}}
(55)&1979&Magnetic monopoles and nonabelian gauge theories. In: {\it Proceedings of the International Conference on Mathematical Physics, Lausanne, Switzerland,1979} (ed. K. Osterwalder), pp.  249-262. Berlin: Springer-Verlag.\\
(56)&1980&Magnetic monopoles. In: {\it Proceedings of the International Conference On High Energy Physics, 1979} (ed. A. Zichichi), pp. 953--957. Geneva: CERN.\\
(57)&1980&Magnetic monopoles and grand unified theories. 
In: {\it Unification of the Fundamental Particle Interactions} (ed. S. Ferrara {\it et al.}), pp. 451--459. New York: Plenum Press.\\
(58)&1981&Classical solutions in gauge theories -- spherically symmetric monopoles -- Lax pairs and Toda lattices. 
In: {\it Current Topics in Elementary Particle Physics} (ed. K. H. M\"utter \& K. Schilling), pp. 199--217. New York: Plenum Press.\\
(59)&1981&Magnetic monopoles and other topological objects in quantum field theory. {\it Phys. Scripta} {\bf 24}, 821--826.\\
(60)&1981&(with P. Goddard) Charge quantization in theories with an adjoint representation Higgs mechanism. \NP {\bf B191}, 511--527.\\
(61)&1981&(with P. Goddard) The magnetic charges of stable self-dual monopoles. \NP {\bf B191}, 528--548.\\
(62)&1982&Self-dual magnetic monopoles. {\it Czech. J. Phys.} {\bf B32}, 529--536.\\
(63)&1982&The structure of self-dual monopoles. 
In: {\it Symposium on Particle Physics: Gauge Theories and Lepton-Hadron Interactions} (ed. Z. Horvath {\it et al.}), pp. 355--395. Budapest, Hungary: Central Research Inst. Physics.\\
(64)&1982&Magnetic monopoles and electromagnetic duality conjectures. 
In: {\it Monopoles in Quantum Field Theory} (ed. N. S. Craigie {\it et al.}), pp. 157--191. Singapore: World Scientific.\\
(65)&1982&(with N. Ganoulis \& P. Goddard) Self-dual monopoles and Toda molecules. \NP {\bf B205}, 601--636.\\
(66)&1982&(with N. Turok) $Z_2$ vortex strings in grand unified theories. \PL {\bf B117}, 193--196.\\
(67)&1983&Relations between grand unified and monopole theories. 
In: {\it Unification of the Fundamental Particle Interactions II} (ed. J. Ellis \& S. Ferrara), pp. 17--28. New York: Plenum Press.\\
(68)&1983&Lectures on gauge theories and lie algebras with some applications to spontaneous symmetry
breaking and integrable dynamical systems. University of Virginia preprint 83-0529 (unpublished). \\
(69)&1983&(with P. C. West) The $N = 4$ supersymmetric $\Er_8$ gauge theory and coset space dimensional reduction. \NP {\bf B217}, 248--284.\\
(70)&1983&(with N. Turok) The symmetries of Dynkin diagrams and the reduction of Toda field equations. \NP {\bf B215}, 470--494.\\
(71)&1983&(with N. Turok) Algebraic structure of Toda systems. \NP {\bf B220}, 491--507.\\
(72)&1984&(with P. Goddard) Algebras, lattices and strings. In: {\it Vertex Operators in Mathematics and Physics} (ed. J. Lepowsky {\it et al.}), pp. 51--96. New York: Springer-Verlag.\\
(73)&1985&(with L. A. Ferreira) Non-compact symmetric spaces and the Toda molecule equations. \CMP {\bf 99}, 365--384.\\
(74)&1985&(with P. Goddard) Kac-Moody algebras, conformal symmetry and critical exponents. \NP {\bf B257}, 226--252.\\
(75)&1985&(with N. Turok) Local conserved densities and zero curvature conditions for Toda lattice field theories. \NP {\bf B257}, 277--301.\\
\end{tabular}

\begin{tabular}{r r p{15cm}}
(76)&1985&(with P. Goddard \& A. Kent) Virasoro algebras and coset space models. \PL {\bf B152}, 88--92.\\
(77)&1985&(with P. Goddard \& A. Schwimmer) The heterotic string and a fermionic construction of the $\Er_8$ Kac-Moody algebra. \PL {\bf B157}, 393--399.\\
(78)&1985&(with P. Goddard \& W. Nahm) Symmetric spaces, Sugawara's energy momentum tensor in two-dimensions and free fermions. \PL {\bf B160}, 111--116.\\
(79)&1985&Kac-Moody algebras: an introduction for physicists. In: {\it Proceedings of the Winter School Geometry and Physics} (ed. Z. Frolík {\it et al.}), pp. 177--198. Palermo: Circolo Matematico di Palermo.\\
(80)&1986&Kac-Moody and Virasoro Algebras in Local Quantum Physics. In: {\it Fundamental Problems of Gauge Field Theory} (ed. G. Velo \&A. S. Wightman), pp. 51--92. New York: Plenum Press.\\
(81)&1986&(with P. Goddard \& A. Kent) Unitary representations of the Virasoro and SuperVirasoro algebras. \CMP {\bf 103}, 105--119.\\
(82)&1986&(with P. Goddard) Kac-Moody and Virasoro algebras in relation to quantum physics. {\it Int. J. Mod. Phys.} {\bf A1}, 303--414.\\
(83)&1986&(with N. Turok) The Toda lattice field theory hierarchies and zero-curvature conditions in Kac-Moody algebras. \NP {\bf B265}, 469--484.\\
(84)&1986&(with P. Goddard) An introduction to Kac-Moody algebras and their physical applications. In: {\it Proceedings of the Santa Barbara workshop on  Unified String Theories} (ed. M. Green \&  D. Gross), pp. 214--243. Singapore: World Scientific.\\
(85)&1986&Infinite dimensional lie algebras and quantum physics. In: {\it Fundamental Aspects of Quantum Theory, Como 1985} (ed. V. Gorrini \& A. Figuereido), pp. 289-293. New York: Plenum Press.\\
(86)&1986&(with P. Goddard, W. Nahm \& A. Schwimmer) Vertex operators for non-simply-laced algebras. \CMP {\bf 107}, 179--212.\\
(87)&1986&(with M. D. Freeman) BRS cohomology in string theory and the no ghost theorem. \PL {\bf B175}, 151--154.\\
(88)&1986&(with M. D. Freeman) The calculation of planar one loop diagrams in string theory using the BRS formalism. \PL {\bf B175}, 155--158.\\
(89)&1987&Lectures on algebras, lattices and strings. In: {\it Supersymmetry, Supergravity, Superstrings '86 } (ed. B. de Wit {\it et al.}), pp. 239--256. Singapore: World Scientific.\\
(90)&1987&Infinite dimensional algebras in modern theoretical physics.  In: {\it Proceedings of the VIIIth International Congress on Mathematical Physics, Marseilles, 1986} (ed. M. Mebkhout \& R. S\'en\'eor), pp. 242-256. Singapore: World Scientific.\\
(91)&1987&(with R. C. Arcuri \& J. F. Gomes) Conformal subalgebras and symmetric spaces. \NP {\bf B103}, 327--339.\\
(92)&1987&(with P. Goddard, W. Nahm, H. Ruegg \& A. Schwimmer) Fermions and octonions. \CMP {\bf 112}, 385--408.\\
(93)&1987&(with P. Goddard \& G. Waterson) Superalgebras, symplectic bosons and the Sugawara construction. \CMP {\bf 112}, 591--611.\\
(94)&1988&The vertex operator construction for non-simply laced Kac-Moody algebras.  In: {\it Infinite-dimensional Lie algebras and their applications, Proceedings, Montreal, 1986} (ed. S. N. Kass), pp. 181-188. Singapore: World Scientific.\\
\end{tabular}

\begin{tabular}{r r p{15cm}}
(95)&1988&Loop algebras, QFT and strings. In:{\it Strings and superstrings: XVIIIth International GIFT Seminar on Theoretical Physics Madrid 1987} (ed. J. R. Mittelbrunn {\it et al.}), pp. 217-285. Singapore: World Scientific.\\
(96)&1988&(ed. with P. Goddard) {\it Kac-Moody and Virasoro algebras: a reprint volume for physicists.} Singapore: World Scientific.\\
(97)&1989&Introduction to string theory: its structure and its uses. {\it Phil. Trans. R. Soc. London} {\bf  A329}, 319-328.\\
(98)&1990&Introduction to conformal invariance and infinite dimensional algebras. In:{\it Physics, geometry, and topology: 1989 Banff NATO ASI} (ed. H.C. Lee), pp. 241-261. New York: Plenum Press.\\
(99)&1991&(with M. -F. Chu, P. Goddard, I. Halliday \& A. Schwimmer) Quantization of the Wess-Zumino-Witten model on a circle. \PL {\bf B266}, 71--81.\\
(100)&1991&(with A. Fring \& H. C. Liao) The Mass spectrum and coupling in affine Toda theories. \PL {\bf B266}, 82--86.\\
(101)&1992&(with A. Fring) The Fusing rule and the scattering matrix of affine Toda theory. \NP {\bf B379}, 429--447.\\
(102)&1993&(with H. C. Liao \& N. Turok) Topological solitons in $\Ar_r$ affine Toda theory. \PL {\bf B298}, 95--102.\\
(103)&1993&(with N. Turok \& J. W. R. Underwood) Solitons and the energy momentum tensor for affine Toda theory. \NP {\bf B401}, 663--697.\\
(104)&1993&(with M. V. Saveliev \& J. W. R. Underwood) On a solitonic specialization for the general solutions of some two-dimensional completely
integrable systems. \PL {\bf B311}, 117--122.\\
(105)&1993&(with N. Turok \& J. W. R. Underwood) Affine Toda solitons and vertex operators. \NP {\bf B409}, 509--546.\\
(106)&1993&(with M. A. C. Kneipp) Crossing and anti-solitons in affine Toda theories. \NP {\bf B408}, 565--573.\\
(107)&1994&(with E. Rabinovici \& A. Schwimmer)  A Class of string backgrounds as a semiclassical limit of WZW models. \PL {\bf B321}, 361--364.\\
(108)&1994&(with A. Fring, P. R. Johnson \& M. A. C. Kneipp) Vertex operators and soliton time delays in affine Toda field theory. \NP {\bf B430}, 597--614.\\
(109)&1995&(with L. Ferreira \& M. V. Saveliev) Orthogonal decomposition of some affine Lie algebras in terms of their Heisenberg subalgebras. {\it Theor. Math. Phys.} {\bf 102}, 10--22.\\
(110)&1996&(with M. A. C. Kneipp) Solitons and vertex operators in twisted affine Toda field theories. \CMP {\bf 177}, 561--582.\\
(111)&1996&Exact electromagnetic duality. {\it Nucl. Phys. Proc. Suppl.} {\bf 45A}, 88--102; {\bf 46}, 1--15.\\
(112)&1997&Lectures on exact electromagnetic duality II. In:{\it Proceedings of the IX Jorge Andr\'e e Swieca Summer School 
Campos de Jord\~ao}  (ed. J.C.A. Barata {\it et al.}), pp. 166--195. Singapore: World Scientific.\\
(113)&1997&Introduction to electromagnetic duality. {\it Nucl. Phys. Proc. Suppl.} {\bf 58}, 43--55.\\
(114)&1998&The monopole. In: {\it Paul Dirac: the man and his work} (ed. P. Goddard), pp. 88-107. Cambridge: Cambridge University Press.\\
(115)&1999&Exact electromagnetic duality. In:{\it Strings, branes and dualities, Cargese, 1997}  (ed. L. Baulieu), pp. 3--31. Berlin: Springer-Verlag.\\
\end{tabular}

\begin{tabular}{r r p{14.8cm}}
(116)&1999&(ed. with P. C. West) {\it Duality and supersymmetric theories}. Cambridge: Cambridge University Press.\\
(117)&1999&(with P. C. West) Particle physics and fundamental theory: introduction and guide to duality and supersymmetric theories.
In: {\it Duality and supersymmetric theories} (ed. D. Olive \& P. C. West), pp. 1--20. Cambridge: Cambridge University Press.\\
(118)&1999&Introduction to duality. In: {\it Duality and supersymmetric theories} (ed. D. Olive \& P. C. West), pp. 62--94. 
Cambridge: Cambridge University Press.\\
(119)&2000&(with M. Alvarez) The Dirac quantization condition for fluxes on four manifolds. \CMP {\bf 210}, 13--28.\\
(120)&2000&Aspects of electromagnetic duality. In: {\it Proceedings, Conference on Nonperturbative quantum effects: Paris, 2000} (ed. D. Bernard  {\it et al.}), {\it JHEP} {\bf PoS tmr2000 021}, pp. 1--7.\\
(121)&2000&Lie algebras, integrability, and particle physics. {\it Theor. Math. Phys.} {\bf 123}, 659--662.\\
(122)&2001&(with M. Alvarez) Spin and Abelian electromagnetic duality on four manifolds. \CMP {\bf 217}, 331--356.\\
(123)&2001&Spin and electromagnetic duality: an outline. Talk at 15th Anniversary Meeting of Dirac Medallists, Trieste, 2000. {\tt hep-th/0104062}.\\
(124)&2001&(with H. Nicolai) The principal SO(1; 2) subalgebra of a hyperbolic Kac-Moody algebra. {\it Lett. Math. Phys.} {\bf 58}, 141--152.\\
(125)&2002&The quantization of charges. {\it Italian Phys. Soc. Proc.} {\bf 79}, 173--183.\\
(126)&2002&(with M. R. Gaberdiel \& P. C. West) A class of Lorentzian Kac-Moody algebras. \NP {\bf B645}, 403--437.\\
(127)&2002&Duality and Lorentzian Kac-Moody algebras. In: {\it Proceedings, International Workshop on Integrable Theories, Solitons and Duality UNESP 2002}  {\it JHEP} {\bf PoS unesp2002  008}, pp. 1--8.\\
(128)&2003&Paul Dirac and the pervasiveness of his thinking. Talk at Dirac Centenary Conference, Waco, Texas, 2002. 	{\tt  hep-th/0304133}.\\
(129)&2003&Charges and fluxes for branes. In: {\it 27th Johns Hopkins Workshop on Current Problems Conference.}  
{\it JHEP} {\bf PoS  jhw2003 019}, pp. 1--7.\\
(130)&2006&(with M. Alvarez) Charges and fluxes in Maxwell theory on compact manifolds with boundary. \CMP {\bf 267} 279--305.\\
(131)&2012&From dual fermion to superstring. In: {\it The Birth of String Theory} (ed. A. Cappelli {\it et al.}), pp. 346--360.\\
\end{tabular}

\end{document}